\documentclass[aps,prd,preprint,eqsecnum,nofootinbib,groupedaddress,floatfix,longbibliography]{revtex4-1}
\pdfoutput=1
\usepackage[utf8]{inputenc}
\usepackage{graphicx,epsf}
\usepackage{amsmath,amsfonts,amssymb}
\usepackage{graphicx}
\usepackage{multirow}
\usepackage{placeins}
\usepackage{wasysym}

\usepackage{hyperref}
\usepackage{xcolor}
\hypersetup{
    colorlinks=true,
    allcolors={blue!60!black}
}

\usepackage[caption=false]{subfig}
\usepackage{todonotes}

\newif\ifdraft
\draftfalse
\newif\ifpreprint
\preprinttrue
\newcommand{\BlackHat}{{\sc BlackHat}}

\newcommand{\SHERPA}{{\sc SHERPA}}

\newcommand{\COMIX}{{\sc COMIX}}

\newcommand{\ntuple}{{$n$-tuple}}

\newcommand{\alpf}{\alpha_f}
\newcommand{\alps}{\alpha_s}

\def\fig#1{fig.~{\ref{#1}}}

\def\figs#1#2{figs.~{\ref{#1}} and {\ref{#2}}}

\def\eqn#1{eq.~(\ref{#1})}

\def\tab#1{table~{\ref{#1}}}

\def\Wjnp1{$W\,\!+\,(n\!+\!1)$}

\def\Wbb{$Wb\bar{b}$}

\def\Wbbp{$W^+b\bar{b}$}
\def\Wbbm{$W^-b\bar{b}$}
\def\Wbbpm{$W^\pm b\bar{b}$}
\def\Wbbj{$Wb\bar{b}+1$-jet}
\def\Wbbjj{$Wb\bar{b}+2$-jet}
\def\Wbbjjj{$Wb\bar{b}+3$-jet}

\def\Wbbn[#1]{$Wb\bar{b}+n$-jet ($n=0,1,2,3$)}
\def\Wbbn{$Wb\bar{b}+n$-jet ($n=0,1,2,3$)}

\def\pT{p_{\rm T}}

\def\HTpartonicp{{\hat H}_{\rm T}'}

\newbox\charbox
\newbox\slabox
\def\s#1{{      % Feynman slash
	\setbox\charbox=\hbox{$#1$}
	\setbox\slabox=\hbox{$/$}
	\dimen\charbox=\ht\slabox
	\advance\dimen\charbox by -\dp\slabox
	\advance\dimen\charbox by -\ht\charbox
	\advance\dimen\charbox by \dp\charbox
	\divide\dimen\charbox by 2
	\raise-\dimen\charbox\hbox to \wd\charbox{\hss/\hss}
	\llap{$#1$}
}}

\begin{document}
\title{
	\ifpreprint
	\hbox{\rm\small
		FR-PHENO-2017-023
	\break}
	\hbox{$\null$\break}
	\fi
        NLO QCD Predictions for $Wb\bar b$ Production in Association with up to Three
	Light Jets at the LHC}

	\author{F. R.~Anger, F.~Febres Cordero, H.~Ita and V.~Sotnikov
	\\
	$\null$
	\\
	Physikalisches Institut, Albert-Ludwigs-Universit\"at Freiburg, D--79104 Freiburg, Germany
}

\begin{abstract}
In this
article we present the next-to-leading order QCD predictions for \Wbbn{}
production at the Large Hadron Collider with $\sqrt{s}=13$ TeV.  We work in the
four-flavor number scheme with a non-vanishing bottom-quark mass and include all
subprocesses at leading electroweak order as well as all heavy-fermion-loop
effects. We show the impact of QCD corrections for total as well as
differential cross sections and make an assessment of theoretical
uncertainties of \Wbb{} production viewed as an irreducible background to $H(\rightarrow
b{\bar b})W$ studies. For the calculations we have
employed an upgraded version of the \BlackHat{}
library which can handle massive fermions in
combination with \SHERPA{}. Our results can be explored
through publicly available \ntuple{} sets.  \end{abstract}

\maketitle

\section{Introduction}

Particle phenomenology at the Large Hadron Collider (LHC) requires the 
measurement of proton-proton collisions with
diverse final states including photons, leptons, heavy and light jets and missing
transverse energy. 
In this article we provide theory predictions for \Wbb{} production in association with up to three light
jets at the LHC in the Standard Model (SM) including next-to-leading
order (NLO) QCD corrections. These processes have rich collider
signatures, including almost all final-state objects mentioned above, and so they provide a
natural test ground for precise measurements of complex signatures at the LHC.
Specifically, they appear as main reducible and irreducible backgrounds
(together with non-resonant top-pair production) to $HW$ associated
production, with the Higgs boson decaying into a bottom-quark pair ($b\bar b$). Currently this process
receives increased attention by both LHC experiments, given
its relevance for constraining the coupling of the Higgs to $b$
quarks. 
Providing results as a function of the light-jet multiplicity is particularly
beneficial, as one expects that universal behavior appears at large
multiplicities~\cite{JetRatios1,JetRatios2,JetRatios3,JetRatios4},
% FFC: removed this sentence here as it doesn't add much. If you add it back,
% please argue why at large multiplicities these resonant contributions do not
% have universal behavior
%
% excluding 
%resonant-top contributions (see e.g. \cite{Denner:2017kzu})
as for
example studied in vector-boson production in association with multiple light
jets in ref.~\cite{BH:Wratios}.  

The first studies of NLO QCD predictions to \Wbb{} production appeared about
twenty years ago~\cite{Ellis:1998fv} and were already included in the first version
of the MCFM program~\cite{mcfm7}, being important in particular for Higgs searches at the Tevatron.
Those initial predictions were performed in the context of the four-flavor number
scheme (4FNS) though they were performed in the approximation of massless quarks
(employing the one-loop helicity amplitudes of ref.~\cite{Bern:1997sc}).
Results including full $b$-mass effects appeared first in
refs.~\cite{FebresCordero:2006sj,Cordero:2009kv} with on-shell $W$ bosons and
subsequent refinements in refs.~\cite{Badger:2010mg,Oleari:2011ey}. NLO QCD
corrections for \Wbbj{} production were obtained
in ref.~\cite{Luisoni:2015mpa,Reina:2011mb}.  Also studies with
more inclusive samples of $b$ jets have been carried out. In
ref.~\cite{Campbell:2006cu} NLO QCD corrections were computed for $W+b+1$-light
jet production, while NLO QCD results were presented for $W$ production in
association with a single $b$ jet in refs.~\cite{Campbell:2008hh,Caola:2011pz}.

Experimental measurements have been carried out by the CDF experiment for $W$
production with one or two $b$ jets~\cite{Aaltonen:2009qi} (including samples
with light jets), by the D0 experiment for inclusive $W$ production with a single $b$
jet~\cite{D0:2012qt}, by the ATLAS experiment for $W$ with up to two $b$
jets~\cite{Aad:2013vka} and also by the CMS experiment for $W$ and two $b$
jets~\cite{Chatrchyan:2013uza,CMS:2016bb}. For a review on these experimental measurements as well as
theoretical predictions of vector boson production in association with $b$ jets
see ref.~\cite{Cordero:2015sba}.

Here we present for the first time NLO QCD corrections to \Wbbjj{} and
\Wbbjjj{} production, and for completeness recompute the cases with zero
and one light jet. 
These ${\cal O}(\alps)$ corrections are computed to the corresponding
leading-order (LO) results at ${\cal O}(\alps^{2+n}\alpf^2)$, with $n$ the numbers
of light jets in the process.
At higher orders in the fine-structure constant $\alpf$, the same signatures can
be obtained from processes involving top quarks which we do not consider here.
Since the early calculations of inclusive \Wbb{}
production~\cite{Ellis:1998fv,FebresCordero:2006sj,Cordero:2009kv} it has been
observed that the NLO QCD corrections are quite large. This is mainly due to
the opening of a gluon-initiated channel as part of the real contributions to
the NLO QCD corrections, as well as to the release of a LO
kinematical constraint which fixes the $p_T$ of the $W$ boson to that of the
$b\bar b$ system. 
%Interestingly these properties are very similar to what is
%observed in NLO QCD corrections to $W$ production in association with a single
%light jet. 
Put differently, those processes suffer from giant
$K$-factors~\cite{Rubin:2010xp}. The corrections to \Wbbj{} production on the
other hand show better behavior~\cite{Luisoni:2015mpa}, and it is expected that
for even larger light-jet multiplicities, the NLO QCD predictions will present more
stable jet-scaling properties. We show that this is indeed the case in our
results.

Attempts to obtain reliable predictions in spite of giant $K$-factors for \Wbb{} production have initially focused on
exclusive analyses~\cite{FebresCordero:2006sj}, including jet vetoes. But sensitivity to the
$p_T^{\rm veto}$ cut tends to spoil the convergence of the perturbative series~\cite{Tackmann:2012bt}.
In this article we employ exclusive sums~\cite{ESums} as an alternative to
stabilize these predictions, considering exclusive combinations of up to two
light jets. The perturbative series for exclusive-sum observables is
better behaved, as confirmed by comparison to LHC
data for $W+1$-jet
production \cite{Aad:2014qxa,ATLAS:ratio2017}. Results based on exclusive sums are expected to 
actually contain some large next-to-next-to-leading order
(NNLO) corrections, and so in the lack of a full NNLO QCD study of \Wbb{} production,
our results represent a useful parton-level prediction for \Wbb{} observables. 
We put particular emphasis on observables associated to $H(\rightarrow b{\bar b})W$ 
production, that is we study the $p_T^{b\bar b}$, $p_T^W$, and
$M_{b\bar b}$ exclusive-sum distributions.

We obtained our results using the \BlackHat{} library~\cite{BlackHatI} after
a significant upgrade and the addition of new algorithms~\cite{BlackHatII}.  Most notably, the new version of the
program allows to compute tree-level and one-loop matrix elements
including massive fermions. This has been achieved by the implementation of the
unitarity method~\cite{Bern:1994zx,Bern:1994cg,Britto:2004nc} extended to massive 
particles~\cite{Bern:1995db},
using a variant of the numerical unitarity
approach~\cite{Ossola:2006us,Ellis:2007br,Ellis:2008ir,Giele:2008ve,BlackHatI} that introduces a modified
spectrum of particles in the loop.  For the real-emission
corrections we use the massive-dipole formalism~\cite{Catani2002}, as
implemented in the \COMIX{} package~\cite{Gleisberg:2008fv} which is part of
the \SHERPA{} Monte Carlo program~\cite{Sherpa}. We store our results in a set
of \ntuple{} files~\cite{BH:Ntuples} which allow fast a-posteriori studies of
our results, including scale variations, parton-distribution functions
(PDFs) and $\alps$ reweighting, and evaluation of additional infrared-safe observables. 

This paper is organized as follows. In section~\ref{sec:calcsetup} we give
our calculational setup,
details of the matrix-element computation
as well as kinematical information, coupling
schemes and other input parameters employed. 
%In section~\ref{sec:bmass}, we
%analyze finite $b$-mass effects by comparing results obtained in both the
%four-flavor number (4FNS) and five-flavor number (5FNS) scheme. 
In
section~\ref{sec:vjets} we present our results for total and differential cross
sections and study their renormalization- and factorization-scale dependence. 
Section~\ref{sec:hw} presents a series of
observables associated to $HW$ production measurements and we show predictions
of exclusive sums 
which combine cross sections of distinct light-jet multiplicities 
and assess
their theoretical uncertainty based on scale dependence, higher-order
contributions and PDF errors. 
We give our
conclusions and outlook in section~\ref{sec:conclusions}.

%%%%%%%%%%%%%%%%%%%%%%%%%%%%%%%%%%%%%%%%%%%%

\section{Calculational setup}
\label{sec:calcsetup}

\begin{figure}[h]
    \centering
    \subfloat[][$qg\rightarrow q^\prime g g W^{\pm}b\bar{b}$]{ \includegraphics[width=0.30\textwidth]{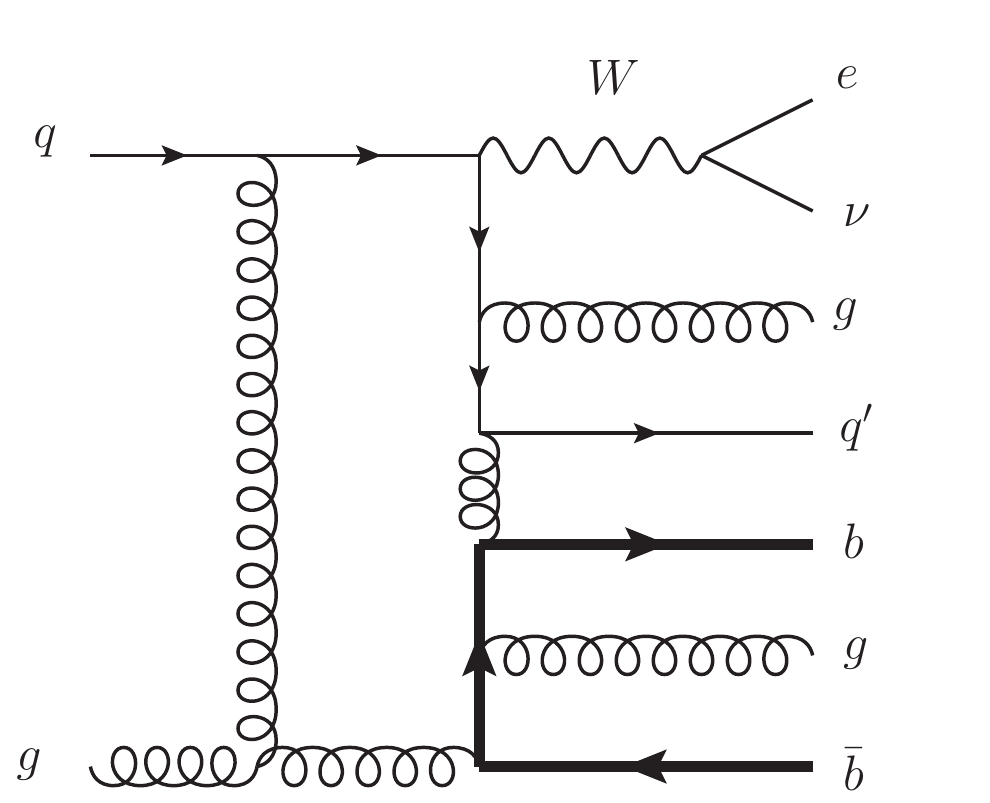}}
    \quad
    \subfloat[][$qg\rightarrow q^\prime g g W^{\pm}b\bar{b}$]{\label{subfloat:nf} \includegraphics[width=0.30\textwidth]{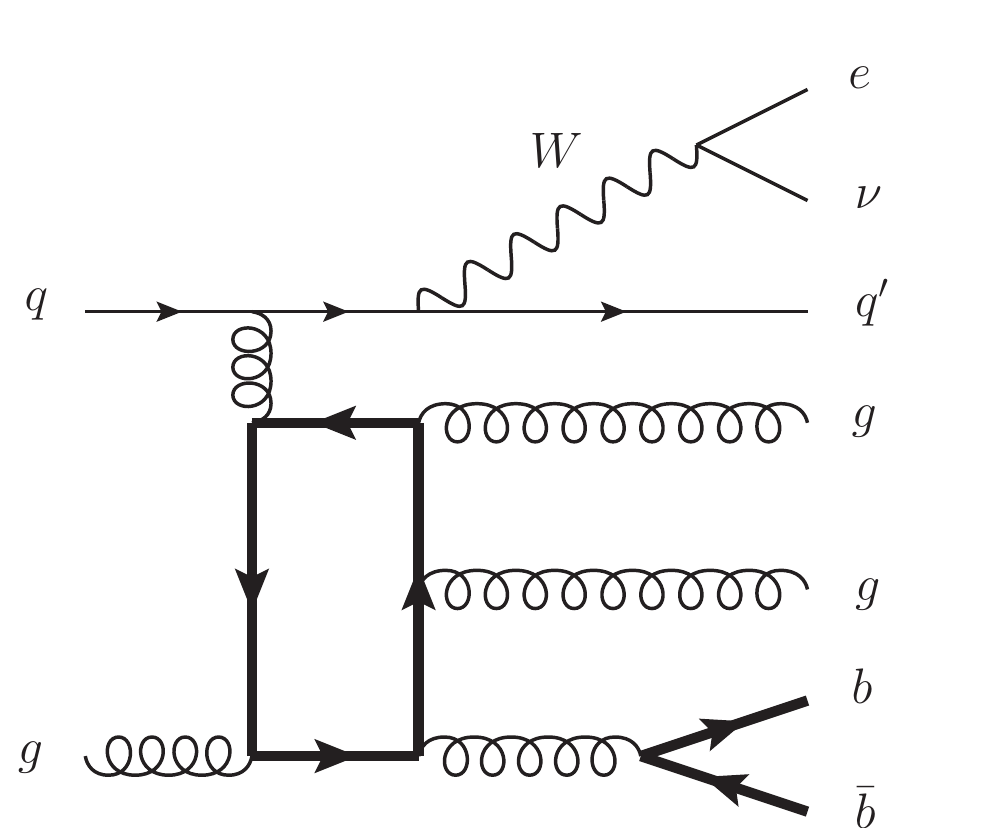}}
    \quad
    \subfloat[][$q{\bar Q}\rightarrow q^\prime \bar{Q} g W^{\pm}b\bar{b}$]{ \includegraphics[width=0.30\textwidth]{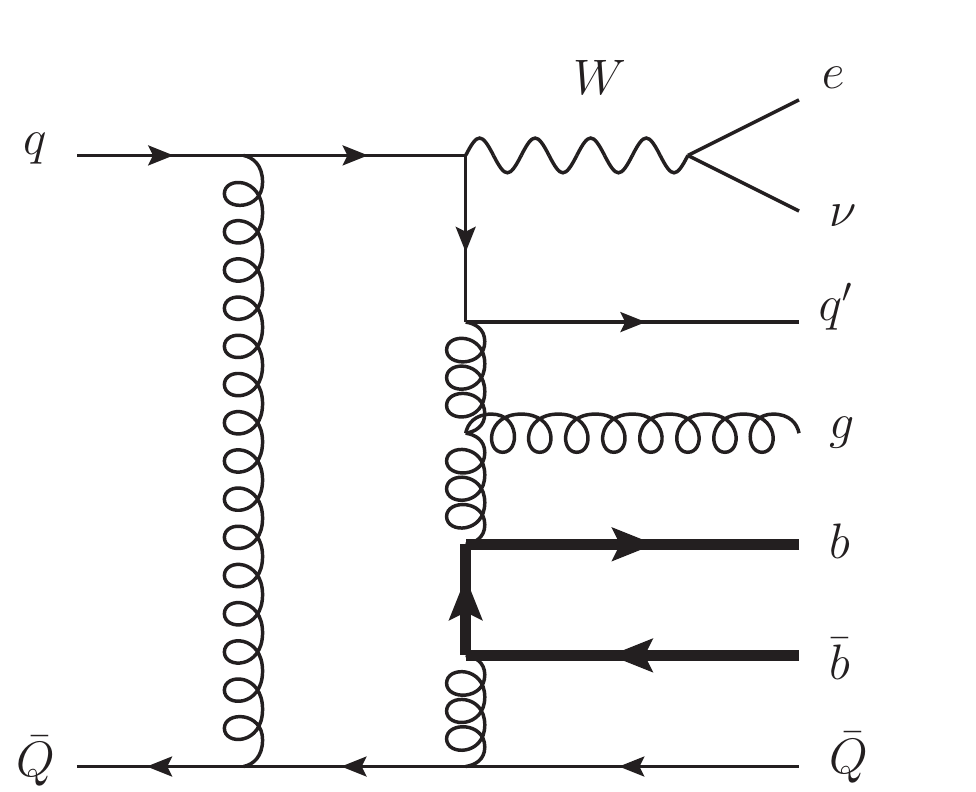}}
    \caption{Representative diagrams for two subprocesses contributing to $pp\rightarrow$ \Wbbjjj{} production.
    The diagram \protect\subref{subfloat:nf} displays a contribution
    from closed loops of top and bottom quarks.}
\label{fig:FDsWbb3j}
\end{figure}

We compute the \Wbbn{} processes in NLO QCD, with the charged vector boson
decayed to a massless charged lepton and its corresponding neutrino. 
The partonic subprocesses of Born and virtual contributions
are obtained from the following subprocesses,
\begin{subequations}
  \begin{align}
    n=0:&\qquad 0\rightarrow Wb{\bar b}q{\bar q}'\ ,\\
    n=1:&\qquad 0\rightarrow Wb{\bar b}q{\bar q}'g\ ,\\
    n=2:&\qquad 0\rightarrow Wb{\bar b}q{\bar q}'gg\ ,\quad  0\rightarrow Wb{\bar b}q{\bar q}'Q{\bar Q}\ ,\\
    n=3:&\qquad 0\rightarrow Wb{\bar b}q{\bar q}'ggg\ ,\quad  0\rightarrow Wb{\bar b}q{\bar q}'Q{\bar Q}g\ ,
  \end{align}
\end{subequations}
by crossing. The labels $q$ and $Q$
denote light-quark flavors and we consider $b$ quarks to be massive (except when
explicitly stated otherwise). Contributions from closed fermions loops of light
quarks as well as top and bottom quarks are included.
Sample Feynman diagrams for the seven-parton amplitudes
are shown in \fig{fig:FDsWbb3j}. The
real-emission subprocesses are obtained from the above list by adding a gluon
or replacing a gluon by a $Q'{\bar Q'}$ pair. 

We obtain fixed-order parton-level predictions and do not include parton-shower
effects.  All observables considered will be constructed from events containing
exactly two observable $b$ jets, defined in an infrared safe
way~\cite{Banfi:2006hf}. We do not introduce any corrections due to possible
mistagging of heavy and/or light jets. 

Below we give more technical details about our results including a brief
description of the one-loop matrix-element computation in the newly setup
\BlackHat{} library, details on the renormalization schemes considered,
numerical stability, validation, Monte-Carlo integration, input parameters, the considered observables as
well our choices for the renormalization and
factorization scales. We end this section with a brief assessment of
$b$ mass effects in our calculation.

\subsection{Virtual Matrix Elements}
\label{sec:VME}

A new version~\cite{BlackHatII} of the \BlackHat{} library~\cite{BlackHatI} is used to compute
the required virtual matrix elements, which includes significant upgrades for
the computation of loop amplitudes with internal and external massive particles.
The library uses the unitarity method~\cite{Bern:1994zx,Bern:1994cg,
Britto:2004nc} and its extension to massive particles~\cite{Bern:1995db} in
order to compute loop amplitudes numerically. These methods have already been
applied by a number of groups for analytic as well as numerical computations
of massive amplitudes (see
e.g.~\cite{Badger:2010mg,Badger:2011yu,Melnikov:2009dn}).
The present implementation is based on the numerical
unitarity approach~\cite{Ossola:2006us,Ellis:2007br,Giele:2008ve,BlackHatI} and its extension to massive quarks~\cite{Ellis:2008ir}. 
In addition, a prescription to map the higher-dimensional
Dirac algebra into four-dimensional objects is used, 
which can be shown to be equivalent to the one given in ref.~\cite{Fazio:2014xea}
with some modifications.

We apply the color decomposition of partial amplitudes in terms of
primitive amplitudes~\cite{Bern:1994fz,Ita:2011ar}. 
Integrands $A(\ell)$ of the primitive amplitudes 
are parameterized as~\cite{Ossola:2006us,Ellis:2007br,Giele:2008ve},
\begin{align}
    \mathcal{A}^{(D_s)}(\ell) =&
  \sum_{i_1<\cdots<i_5} \dfrac{{\bar{e}}^{(D_s)}_{i_1i_2i_3i_4i_5}(\ell)}{d_{i_1}d_{i_2}d_{i_3}d_{i_4}d_{i_5}}~+~
  \sum_{i_1<i_2<i_3<i_4} \dfrac{{\bar{d}}^{(D_s)}_{i_1i_2i_3i_4}(\ell)}{d_{i_1}d_{i_2}d_{i_3}d_{i_4}}\nonumber\\
  &~+~\sum_{i_1<i_2<i_3} \dfrac{{\bar{c}}^{(D_s)}_{i_1i_2i_3}(\ell)}{d_{i_1}d_{i_2}d_{i_3}}~+~
  \sum_{i_1<i_2} \dfrac{{\bar{b}}^{(D_s)}_{i_1i_2}(\ell)}{d_{i_1}d_{i_2}}~+~
  \sum_{i_1} \dfrac{{\bar{a}}^{(D_s)}_{i_1}(\ell)}{d_{i_1}}\ ,
  \label{eq:master}
\end{align}
where $d_{i_j}$ are inverse propagators, $\ell$ is the $D$-dimensional loop momentum,
$D_s\ge D$ is the dimension of spin states of the loop
particles, and the sums run over all inequivalent products of inverse propagators.
If all external states are four dimensional, terms with at most five
propagator denominators are irreducible. The numerators ${\bar{e}(\ell)}$,
${\bar{d}(\ell)}$, ${\bar{c}(\ell)}$, ${\bar{b}(\ell)}$ and ${\bar{a}(\ell)}$ are polynomials in the loop
momentum and can be decomposed into surface terms, which vanish upon integration
over the loop momentum, and master terms, which are associated to master
integrals. We customarily choose these terms such that the set of master integrals
is formed by 1-, 2-, 3-, and 4-point scalar integrals, as well as by ``extra'' integrals originating from
squares of $(D-4)$-dimensional parts of the loop momentum~\cite{Giele:2008ve,Ossola:2008xq}.

Unitarity methods obtain the above numerators directly from on-shell
tree amplitudes, thus avoiding an explicit reduction of tensor and higher-rank
integrals, as in the associated OPP reduction method~\cite{Ossola:2006us}.
Suitable loop-momentum parameterizations allow to set $n\leq5$ loop propagators
on-shell. On these on-shell phase spaces the amplitude factorizes into a product
of $n$ tree amplitudes such that in these limits the integrands in
\eqn{eq:master} can be computed from simple tree-level input.

We have implemented loop-momentum parameterizations for on-shell conditions
associated to arbitrary (real or complex) masses.  We compute the required
tree amplitudes numerically via Berends-Giele (BG) off-shell
recurrence relations~\cite{Berends:1987me}. In particular, this allows for an
efficient and flexible tree generation for both complex and $D$-dimensional
momenta. Special attention needs to be paid to the computation of single cuts
for the extraction of tadpole coefficients
and double cuts for the extraction of coefficients of bubbles with a single
on-shell leg. When computing the required single
and double cuts, explicit divergences associated to self-energy insertions on external legs
are encountered. These contributions are removed and accounted for by mass 
and wave-function renormalization.
In order to do this, we follow
ref.~\cite{Ellis:2008ir} and adjust the tree-diagram generation
to remove these contributions from the cuts. Alternative approaches have been presented in
ref.~\cite{Britto:2011cr} and, recently, in ref.~\cite{Badger:2017gta}.

We compute intermediate expressions in the four-dimensional helicity (FDH)
scheme. The limit $D_s\to 4$ in
\eqn{eq:master} has thus to be taken after the Dirac and vector algebra
are evaluated. 
We address this with an approach based on $D_s$-dimensional unitarity cuts~\cite{Giele:2008ve}.
Whenever
possible we reduce the $D_s$-dimensional algebra and states to lower dimensions.
This is in close analogy to the
known decomposition of the $D_s=6$ dimensional gluon amplitude to a $D_s=5$
amplitude plus a scalar contribution~\cite{Bern:1994cg}. By doing so, we avoid
an overhead of numerical computations in higher-dimensional representations.
Our implementation can be equivalently derived from the four-dimensional
(re-)formulation of FDH~\cite{Fazio:2014xea} with some modifications. Details
will be discussed elsewhere \cite{BlackHatII}.

Finally, we implemented the one-loop scalar integrals
based on refs.~\cite{tHooft:1978jhc,Ellis:2007qk,Denner:2010tr}. The integrals are setup for evaluation
in higher-precision arithmetics and have been
systematically checked by comparing against the {\sc OneLOop} Fortran
library~\cite{vanHameren:2010cp}, as well as against the massless
one-loop scalar integrals implemented in earlier versions of \BlackHat{}.

\subsection{Renormalization}
We use the FDH variant of dimensional regularization in intermediate steps to
regularize UV and IR divergences. At the end we convert the renormalized
amplitude to the 't Hooft-Veltman (HV) scheme~\cite{tHooft:1972tcz}.

In \tab{tab:renorm} we summarize the renormalization counterterms
in the FDH scheme. For all external states we use the on-shell
scheme and for the QCD coupling we use the $\overline{MS}$ scheme. The gluon
wave-function-renormalization counterterm receives contributions from all
active heavy quarks. In the decoupling limit this counterterm is set to
zero.
\begin{table}[h]
  \centering
  \begin{ruledtabular}
    \begin{tabular}{lcll}
      \textbf{Renormalization} & \textbf{Scheme} & \textbf{Counterterm} \\
      \noalign{\vskip 2mm}
      \colrule
      \noalign{\vskip 2mm}
      %\rule{0pt}{1ex}\\
    %\small
      %\toprule
      Heavy quark wave function   & on-shell & $\displaystyle \delta_{2,i} ~=~ \frac{N_c^2-1}{2N_c} \left( \frac{3}{\epsilon} + 5 + 3 \ln{\frac{\mu^2}{m_i^2}} \right)$\\
      Light quark wave function   & on-shell & 0\qquad(UV+IR cancellation) \\
      Quark mass            & on-shell & $\displaystyle \delta_{m_i} ~=~ \delta_{2,i}\quad\text{}$\\
      Gluon wave function   & on-shell & $\displaystyle \delta_3 ~=~ \frac{3}{\epsilon} + \sum_i \frac{1}{3}\ln{\frac{\mu^2}{m_i^2}}$\\
      QCD coupling & $\overline{MS}$ & $\displaystyle \delta_{\alps} ~=~ \frac{1}{\epsilon} \left( \frac{11}{3}N_c - \frac{2}{3}(N_f+N_h) \right) - \frac{N_c}{3}$\\
      %\midrule
      \noalign{\vskip 2mm}
      \hline
      \noalign{\vskip 2mm}
      Decoupling shift & --- & $\displaystyle     \Delta_i ~=~  -\frac{2}{3}\ln{\frac{\mu^2}{m_i^2}} $\\
      %\botrule
    \end{tabular}
  \end{ruledtabular}
  \caption{Renormalization counterterms. Here $\mu$ is the renormalization
  scale, $m_{i}$ are the masses for heavy quarks, $N_f$ is the number of light flavors,
  $N_h$ the number of heavy flavors, and $N_c$ the number of colors.
  A common factor of $-4\pi\alpha_s c_\Gamma$ has been factored out.
}
  \label{tab:renorm}
\end{table}
In the 4FNS, the number of light flavors $N_f$ is set to four, and the
number of heavy flavors $N_h$ is set to two, accounting for bottom-
and top-quark loops. For each heavy quark included we also add a finite
decoupling shift. Thus the full renormalized amplitude $\mathcal{A}^{(ren)}$ is
obtained from an amplitude with renormalized quark masses, $\mathcal{A}^{(bare)}_{m_R}$ as
\begin{equation}
  \mathcal{A}^{(ren)} =
  \mathcal{A}^{(bare)}_{m_R} - 4\pi \alpha_s c_\Gamma 
  \left( \sum_i N_{Q_i} \frac{\delta_{2,i}}{2} + N_{g}\delta_3 + \frac{N_{\alps}}{2}\left(\delta_{\alps} + \sum_{i\in N_h}\Delta_i\right) \right)  \mathcal{A}^{(born)},
  \label{renfull}
\end{equation}
where $\displaystyle c_\Gamma={(4\pi)^{-(2-\epsilon)}{\Gamma(1+\epsilon)\Gamma^2(1-\epsilon)}/\Gamma(1-2\epsilon)}$,
$N_{Q_i}$ is the number of external heavy quarks of flavor $i$, $N_g$ is the number of external
gluons and $N_{\alps}$ is the $\alps$ order (at Born level).

The renormalization of quark masses cannot be represented as a contribution proportional to the tree amplitude. We explicitly
compute mass counterterm contributions using a dedicated recursive tree-like computation at the level of primitive loop amplitudes.

The conversion to the 't Hooft-Veltman scheme is performed by a finite shift~\cite{Signer:2008va}
\begin{equation}
  \mathcal{A}^{(ren)}_{HV} - \mathcal{A}^{(ren)}_{FDH} = -4\pi \alpha_s c_\Gamma\left(N_{g}~\frac{N_c}{6} + \frac{N_q}{4}\left(N_c -\frac{1}{N_c}\right)\right)\mathcal{A}^{(born)},
  \label{schemeshift}
\end{equation}
where $N_q$ stands for the number of light external quarks in the respective amplitude.

\subsection{Numerical Stability}
The numerical unitarity approach has proven to be numerically stable even when
computing high-multiplicity one-loop matrix elements (see for example~\cite{BH:W3jDistributions,BH:W5j}). Nevertheless,
there are phase-space regions where the on-shell loop-momentum parameterizations break down. This occurs particularly where Gram determinants are close
to zero. For such configurations incomplete cancellations of large contributions can
lead to a loss of numerical precision. 
In this section we study
the numerical stability of our  massive one-loop matrix elements by comparing
results computed in normal-production mode ($d\sigma_V^{\rm prod}$) with computations performed in
quadruple precision~\cite{QD} ($d\sigma_V^{\rm HP}$), i.e. up to 32 digits of precision.

\begin{figure}[h]
        \centering
        \includegraphics[scale=1.1]{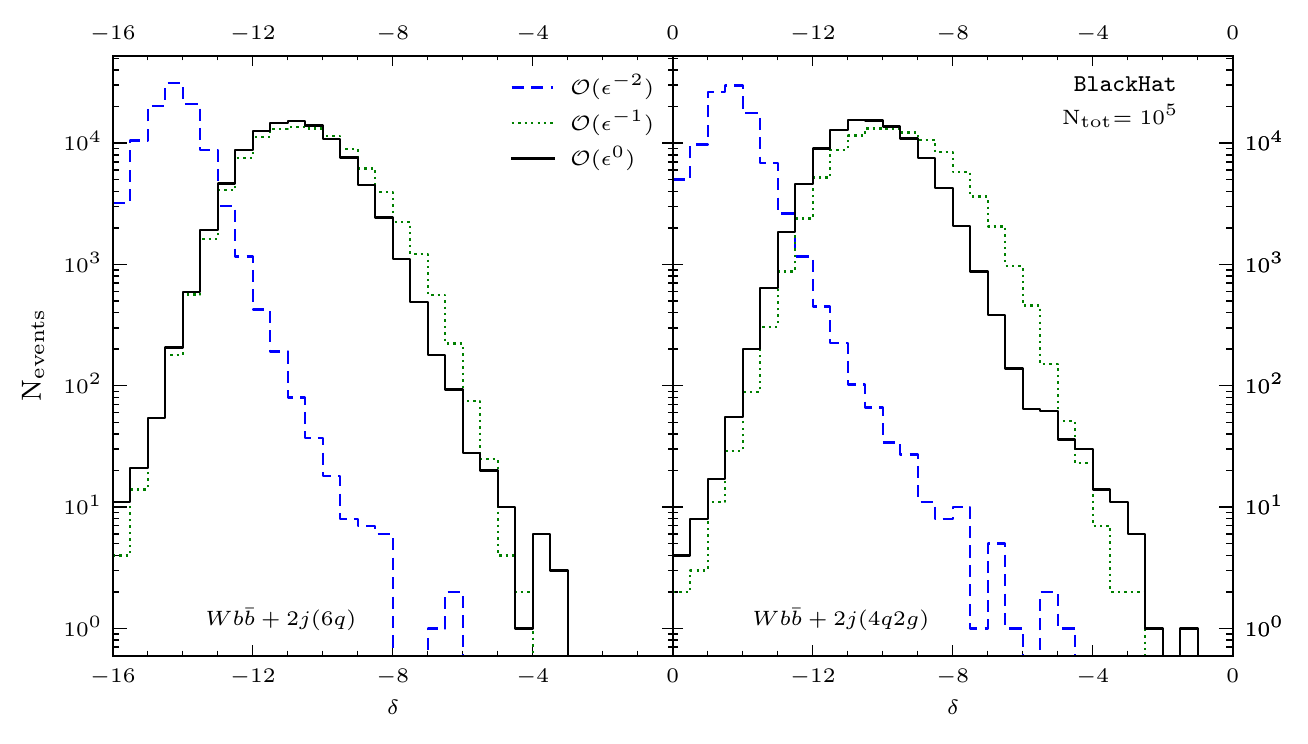}
    \caption{
      The logarithmic relative error of the full-color matrix elements
for two types of subprocesses contributing to the \Wbbjj~production calculation. On the
left we show results for the
six-quark and on the right for four-quark matrix elements, respectively.
We use a set of $10^5$ phase-space points
sampled for the LHC setup with $\sqrt{s}=13$ TeV, in the same way as the phenomenological
study presented in this paper, and use a dedicated calculation in
quadruple precision for computing reference results.
The dashed (blue) line represents the precision of the double pole, the dotted
(green) line represents the single pole and the
solid (black) line the precision of the finite piece of the calculation.}
    \label{fig:stabilityWbb2j}
\end{figure}
\begin{figure}[h]
        \centering
        \includegraphics[scale=1.1]{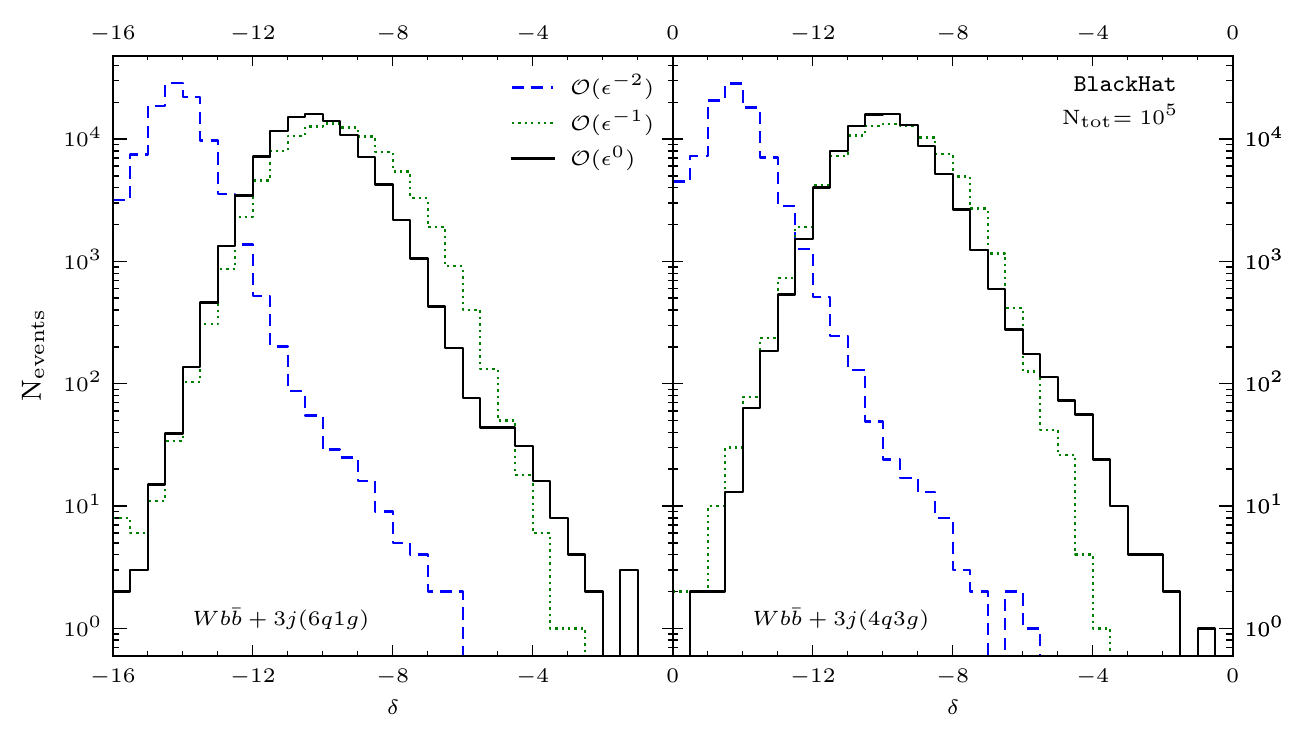}
    \caption{As in \fig{fig:stabilityWbb2j} but for \Wbbjjj{} production,
considering only the leading-color contributions to the one-loop matrix elements.
    On the left we show results associated to the six-quark and on the right the ones associated to
    four-quark matrix elements.}
    \label{fig:stabilityWbb3j}
\end{figure}

We produce histograms of the logarithmic relative error $\delta$:
\begin{equation}
  \delta = \log_{10}\left(\frac{\left|d\sigma^{\text{prod}}_V - d\sigma^{\text{HP}}_V\right|}{\left|d\sigma^{\text{HP}}_V\right|}\right)\ ,
  \label{reldiff}
\end{equation}
for sample subprocesses of the most complex types in our calculations. 
The superscripts ``prod'' and ``HP'' mean normal-production evaluation and
quadruple floating-point evaluation respectively.
In \figs{fig:stabilityWbb2j}{fig:stabilityWbb3j} we show a numerical stability plot for \Wbbjj{} and
\Wbbjjj{} production respectively. For the \Wbbjj{} process we include the two types of
subprocesses: those associated to the four-quark ($0\rightarrow Wb{\bar b}q{\bar q}'gg$)
and the six-quark ($0\rightarrow Wb{\bar b}q{\bar q}'Q{\bar Q}$) matrix elements, respectively.
Similarly, for the \Wbbjjj{} process we show results for the four-quark ($0\rightarrow Wb{\bar b}q{\bar q}'ggg$)
and six-quark ($0\rightarrow Wb{\bar b}q{\bar q}'Q{\bar Q}g$) matrix elements.
Each figure shows the estimated precision of the double-pole and
single-pole terms as well as the finite part in the $\epsilon$ expansion of the 
matrix elements. The double-pole term is
commonly computed with an accuracy of 14 digits, while the single-pole and the
finite-part distributions peak at about 10 digits.

To control the precision of the computation we implemented a rescue system which 
identifies phase-space points 
which lead to numerically unreliable computations
by performing checks at several
stages of the calculation\footnote{With some minor adaptations we use the same
implementation as described in \cite{BlackHatI,BH:W3jDistributions}.}. Whenever
any of those checks fail we switch the computation to use higher-precision 
arithmetics  locally, i.e. only for parts of the computation which failed the check.
Typically, the fraction of time spent on these recomputations is small.

Overall we observe
very good precision, with less than $1$ in $10^4$ phase-space
points computed to an accuracy worse than three digits\footnote{This precision target 
is chosen to allow for matrix element evaluation to a much better
precision than the achievable statistical 
error in Monte Carlo integrations.}.
Some of the few points in the low-precision tail
%of the histogram are caused by instabilities in finite box
%integrals with three and four internal masses as well as by an accidental failure of the rescue system.
%Most of the time instabilities 
can be traced
to the fact, that the scale set by the bottom mass $m_b$ and other scales of the problem $s_{ij}$ at
sufficiently high energy are separated by several orders of magnitude.
Ratios of the form $(m_b^2/s_{ij})^k$ enter loop-momentum parameterizations and can cause loss of precision.
We have observed that
the contribution of the points in the low-precision tail is several orders of magnitude smaller than the total
statistical errors for the observables studied.

\subsection{Validation}
We have carried out a number of checks with the new implementation of \BlackHat{}~\cite{BlackHatII}.
In particular, we have systematically reproduced all massless results
that were carried out with all the earlier versions of the library, and found
excellent agreement.

We also apply a number of internal checks. For each primitive amplitude we
check consistency of the integrand reduction by attempting to compute a
high-rank tensor coefficient, which is known to vanish from power counting. We
also check that the matrix elements reproduce the known IR and UV singularity structure~\cite{Catani:2000ef}. Both
of these checks are also used to control the precision of the computation.
Explicit cancellation of the infrared poles of the renormalized one-loop matrix
elements is also checked by comparing to the integrated subtraction terms, after
PDF renormalization, as performed with the~\SHERPA{} library.

We also reproduced the results presented by Ellis \textit{et al.}
in~\cite{Ellis:2008ir} at the level of primitive amplitudes. Even more, we have
cross checked fully interfered, matrix-element squared results against
publicly available matrix-element generators. This includes comparisons of all
one-loop matrix elements necessary for $pp\rightarrow t\bar t+(\leq 2)-$jet and
$pp\rightarrow b\bar b+(\leq 2)-$jet at NLO QCD against the
{\sc Recola}~\cite{Actis:2016mpe} and
{\sc OpenLoops}~\cite{Cascioli:2011va} libraries (both using the
{\sc Collier} library~\cite{Denner:2016kdg}), as well as all the ones needed
for $Wb\bar b+(\leq3)-$jet with {\sc Recola}.

Finally, we made dedicated comparisons at the level of physical observables
against the MCFM program~\cite{mcfm7} for the inclusive production of \Wbb{}
at NLO QCD at the LHC with $\sqrt{s}=13$ TeV. Agreement was found at the permil level for total cross sections
and differential distributions.

\subsection{Monte Carlo Integration}
%\subsection{Real-emission corrections}

NLO QCD corrections require the integration of Born, virtual and real-radiation
matrix elements.  We provide virtual matrix elements to the \SHERPA{} 
Monte Carlo program~\cite{Sherpa}, which are then combined with Born and 
real-radiation matrix elements and integrated consistently.

The results in this article were obtained by exploiting the color structure of the matrix
elements to increase the efficiency of the phase-space integration.  To this
end we split up the virtual matrix elements into leading and subleading color
terms~\cite{BH:W3jDistributions,Ita:2011ar} and sample them independently over
phase space. Given the color suppression of the subleading color terms these
computationally expensive contributions need to be evaluated less often to reach a
given total statistical-integration error.

The infrared singularities of the real-emission corrections have to be cancelled explicitly against their counterparts in the virtual matrix elements before numerical integration. We
use the subtraction scheme based on massive dipoles~\cite{Catani2002} in our
calculation.  In particular, we employ the automated implementation in the \COMIX{} package~\cite{Gleisberg:2008fv}, which is part of the \SHERPA{}
library~\cite{Sherpa}. The latter implementation has been checked extensively
in the recent calculation of $t\bar t$ production in association
with up to three light jets~\cite{Hoche:2016elu}. 

We also employ the \SHERPA{} library to handle the subprocess generation and
mapping. Furthermore, we rely on \SHERPA{} for the phase-space integration and 
use its internal Analysis package. 
We compute fixed-order parton-level predictions and
include neither parton-shower effects nor hadronization corrections or other
non-perturbative effects.
We store our results in flexible
\texttt{ROOT}~\cite{ROOT} \ntuple{} files, which allow for a-posteriori variations of the strong coupling,
PDF sets and choices of renormalization and factorization
scales of the NLO QCD results. The format was developed in ref.~\cite{BH:Ntuples}.

\subsection{Input Parameters: Partons Distributions, Couplings and Masses}
\label{sec:base_setup}

We take parton-distribution functions (PDFs) from {\tt CT14}~\cite{CT14},
with LO ({\tt CT14llo\_NF4}) and NLO ({\tt CT14nlo\_NF4}) PDF sets, as
implemented in the LHAPDF library~\cite{LHAPDF}. The strong
coupling is fixed accordingly, with $\alps(M_Z)=0.125$ at LO and
$\alps(M_Z)=0.1128$ at NLO. We evolve $\alps(\mu)$ with
the QCD beta function for four massless quark flavors, for all $\mu$. This is
achieved by introducing a decoupling shift for massive quarks (see above). We
use a one-loop running of $\alps$ at LO and a two-loop running at NLO. We
choose $m_b=4.75$ GeV, consistent with the PDF sets, and set the top-quark
mass in closed-loop contributions to $m_t=172$ GeV.

We work at leading order in the electroweak coupling and fix the
$W$-boson couplings to fermions with the SM input
parameters as specified in \tab{tab:ewinput}. We use the $G_\mu$
scheme \cite{Denner2000c} and compute the parameters $\alpf(M_Z)$,
$\sin^2(\theta_W)$ and $g_W^2$ using the tree-level relations
\begin{align}
\sin^2(\theta_W) &= \left(1-\frac{M_W^2}{M_Z^2}\right)\ , & \alpf(M_Z)&=\frac{\sqrt{2}}{\pi}G_F M_W^2
  \sin^2(\theta_W)\ ,\notag\\
g_W^2&=\frac{4\pi\alpf(M_Z) }{\sin^2(\theta_W)}\ .
\end{align}

\begin{table}[]
\centering
\caption{Electroweak parameters used in this work. We use the $G_\mu$
  scheme with real parameters and leading-order relations. The input parameters
are chosen in accordance with 2016 PDG values~\cite{Patrignani:2016xqp}.}
\label{tab:ewinput}
  \begin{tabular}{p{3.5cm}p{5cm}}
    \hline\hline
    \noalign{\vskip 2mm}
    Parameter & Value  \\
    \noalign{\vskip 2mm}
    \colrule
    \noalign{\vskip 2mm}
    $G_F$ & $1.1663787 \times 10^{-5}$ GeV$^{-2}$ \\
    $M_W^{\text{OS}}$& $80.385$ GeV \\
    $M_Z^{\text{OS}}$& $91.1876$ GeV \\
    $\Gamma_W$& $2.085$ GeV \\
    $\alpf(M_Z)$ & $1/132.23$ (calculated)\\
    $\sin^2(\theta_W)$ & $0.22290$ (calculated)\\
    $g_W^2$ & $0.42635$ (calculated)\\
    \hline\hline
  \end{tabular}
\end{table}

The lepton-pair invariant mass follows a relativistic Breit-Wigner distribution,
with the mass and width as specified in~\tab{tab:ewinput}.
We approximate the Cabibbo-Kobayash-Maskawa (CKM) matrix by a unit matrix. This
results in a small change of the total cross sections for the setup we use, as
estimated by LO evaluations with the full CKM matrix.
Indeed we find that these differences are of the order 1\% for \Wbb{} production and
below 0.5\% for \Wbbj{} and \Wbbjj{} production.

All light quarks ($u$, $d$, $s$, $c$) are treated as massless. We do
include contributions from virtual bottom and top quarks, and we confirm the
expected percent-level effect on
cross-sections~\cite{BH:W4j,BH:Z4j,Campbell:2016tcu}. We work with a single massless
lepton pair, an approximation that can be applied to the electron or muon
families.

\subsection{Kinematics, Observables and Exclusive Sums}
\label{sec:kin}
For completeness, we state the definitions of the standard observables
used in our analysis. The pseudorapidity $\eta$ is given by
\begin{align}
  \eta = -\ln\left(\tan\frac{\theta}{2}\right),
\end{align}
with the polar angle $\theta$ with respect to the beam axis. The
angular separation between any two objects (partons, jets or
leptons) is denoted by
\begin{align}
  \Delta R = \sqrt{(\Delta \phi)^2+(\Delta \eta)^2},
\end{align}
where $\Delta\phi$ is the difference in the azimuthal angle in the
plane transverse to the beam axis and
$\Delta\eta$ the difference in the pseudorapidities. 
The total partonic transverse energy is given by
\begin{equation}\label{eq:htpart}
  \HTpartonicp=\sum_j p_{\rm T}^j+E_{\rm T}^W,
\end{equation}
where the sum runs over all final state partons $j$, independent of
whether or not they are inside a jet that passes the cuts. The scalar
transverse momentum $p_T$ of a parton is given by
$p_T=\sqrt{p_x^2+p_y^2}$ and the transverse
energy of the $W$ boson is computed from
\begin{align}
  E_T^W=\sqrt{M_W^2+\left(p_T^W \right)^2}\ .
\end{align}
The total partonic transverse energy is not directly measurable. Nevertheless, it is
a suitable candidate for choosing unphysical renormalization and
factorization scales since changing the cuts does not affect the
matrix element at a given phase-space point. Jet invariant masses are
defined by
\begin{align}
  M_{ij}^2 = \left(p_i^{\text{jet}}+p_j^{\text{jet}}\right)^2,
\end{align}
where jets are labeled in order of decreasing transverse momentum
$p_T$. The transverse mass of the $W$ boson is computed from the
kinematics of its decay products
\begin{align}
  M_T^W=\sqrt{2E_T^eE_T^\nu(1-\cos(\Delta\phi_{e\nu}))}\ .
\end{align}

In section~\ref{sec:hw}, we show predictions based on exclusive
sums~\cite{ESums}, which combine exclusive ($\sigma^{\text{exc}}_{n}$) and
an inclusive prediction ($\sigma^{\text{inc}}_{n}$) for distinct light-jet
multiplicities $n$.
We study
exclusive sums for observables in \Wbb{} production (0 light jets) and use our high-multiplicity
results to define the exclusive sums labeled `NLO+' and `NLO++',
\begin{align}\label{eq:excsums}
  \sigma^{\text{NLO+}}_0 &= \sigma^{\text{exc}}_0 + \sigma^{\text{inc}}_1\ , &
\sigma^{\text{NLO++}}_0 &= \sigma^{\text{exc}}_0 +\sigma^{\text{exc}}_1+
\sigma^{\text{inc}}_2\ .
\end{align}

\subsection{Dynamical Scale Choice}
\label{sec:scale}

We present renormalization and factorization scale dependence of cross sections using correlated variations of a central scale choice $\mu_0$ by factors of
$(1/2,1/\sqrt{2},1,\sqrt{2},2)$. We have also explored independent variations of
those scales and find similar results as with the correlated variations. 
We choose $\HTpartonicp/2$ (\ref{eq:htpart})  as the functional form for the
central scale $\mu_0$ and keep factorization and renormalization
scales equal, $\mu_R=\mu_F=\mu_0$. The scale $\mu_0=\HTpartonicp/2$ has proven to be a sensible
choice as it tends to reduce shape changes and the global size of quantum corrections
when going from leading to next-to-leading order (see for example~\cite{BH:W5j,BH:Z4j,BH:Wratios}).
In general, NLO corrections are less sensitive to the choices of scale, as long
as the scale reflects the hardness of the Born process~\cite{BH:W3jDistributions}.
We will label results for leading-order QCD with the central scale
$\mu_0=\HTpartonicp/2$ by `LO' and the corresponding results at next-to-leading
order QCD by `NLO'.

\subsection{Effects of a Finite $b$ Mass}
\label{sec:bmass}
Mass effects in \Wbb{} have been studied since the early NLO QCD
calculations in ref.~\cite{FebresCordero:2006sj}. They are expected to be small when two
well-defined $b$ jets are considered, and ratios of $m_b^2$ to typical invariants
are small. Nevertheless, their contributions are fundamental when studying
inclusive $b$-jet production at hadron colliders (see for example 
refs.~\cite{Campbell:2008hh,Caola:2011pz}).

%mbb spectrum Wbb & Wbbj ratio massive/massless
%%%%%%%%%%%%% FIGURE %%%%%%%%%%%%%%%%%%
\begin{figure}[ht]
\centering
%\begin{centering}
\includegraphics[clip,scale=1.0]{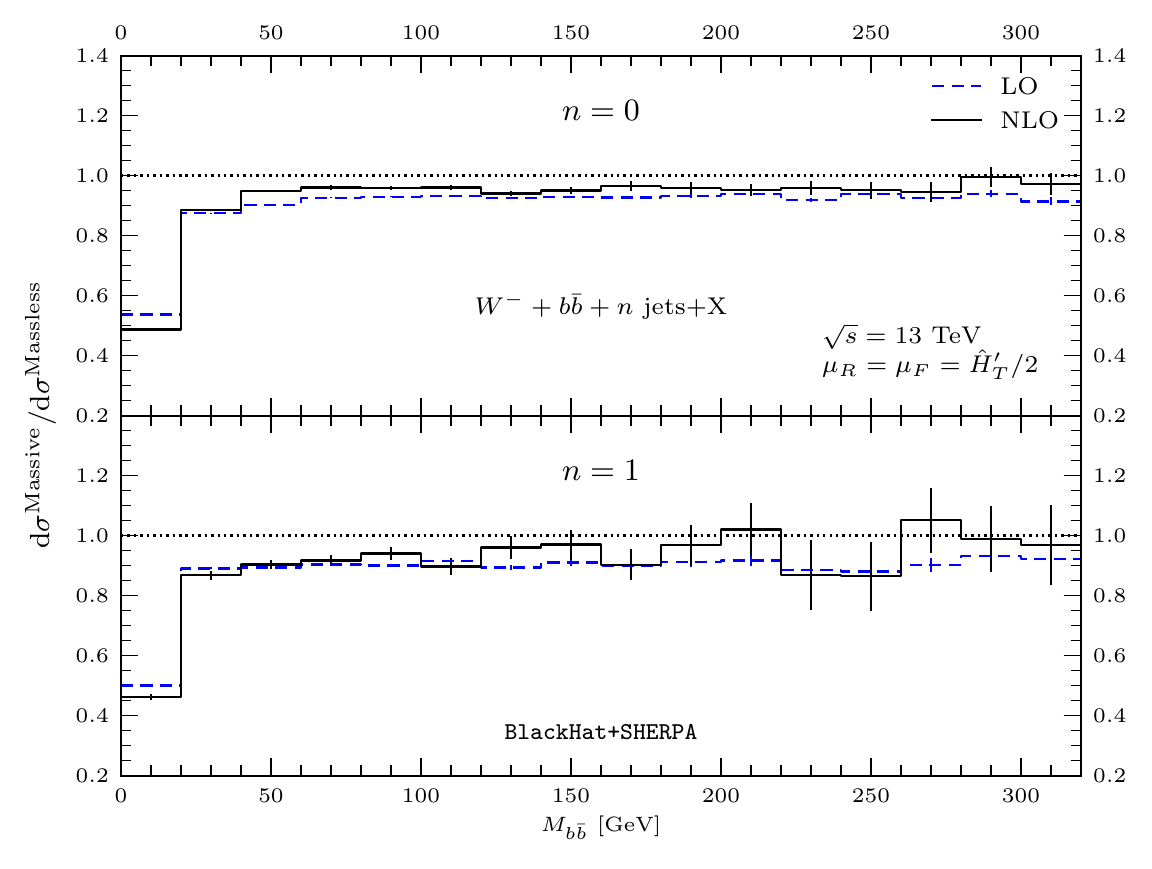}
%\end{centering}
  \caption{Ratio of the invariant mass spectrum of the $b\bar b$ system for 4FNS
    result to the 5FNS ones, for \Wbbm{} (top) and \Wbbm+1-jet (bottom) production.
    The ratios are taken at LO (dashed blue line) and at NLO (solid black line).
    Statistical errors are shown as thin vertical lines. We include a dotted
    horizontal line at a ratio value of 1.}
  \label{fig:ratWmmbb}
\end{figure}
%%%%%%%%%%%%%%%%%%%%%%%%%%%%%%%%%%%%%%%

In order to highlight these effects, in \fig{fig:ratWmmbb} we show the ratio of
a computation performed in the 4FNS consistently keeping the mass of the $b$
quarks, to that of a corresponding massless calculation performed with massless
$b$ quarks in the five-flavor number scheme (5FNS). In the latter we use the
PDFs from CT14~\cite{CT14}, denoted by \texttt{CT14llo} at LO and
\texttt{CT14nlo} at NLO.  We notice that for values of $M_{b\bar b}$ above 50
GeV, the ratios stabilize rapidly at about 0.95 for \Wbb{} production and at
0.9 for \Wbbj{} production, while for values below we have a strong deviation
with the massless calculation more than doubling the massive one. This is to be
expected as phase space constrains the production of massive
$b$ quarks in these regions and also $m_b^2/M_{b\bar{b}}^2$ terms in
the matrix elements can be important.

The mass effects are stable with respect to quantum corrections, as we can
deduce from the similarity of the LO and NLO ratios. We notice that the deviation
from 1 at large $M_{b\bar b}$ is smaller than the scale-dependence bands of the
NLO results, which can be used as a proxy of unaccounted higher-order
corrections. 

The observed behavior is very similar at what was studied in
ref.~\cite{FebresCordero:2006sj} in the case of \Wbb{} production, and here we
extend it to \Wbbj{} production. It is important to mention that although the
computations in \fig{fig:ratWmmbb} are consistent results in the 4FNS and in the
5FNS, they have the same diagrammatic content at Born level, and in
particular there is no $b$-initiated subprocess in the massless
calculation. For
that reason the comparison presented is attributable to $b$-mass effects in the
matrix elements and in phase-space generation, together with the corresponding
differences from the PDFs and their corresponding running couplings. A more
systematic 4FNS vs. 5FNS comparison including \Wbbjj{} and \Wbbjjj{}
production would be more relevant to compare the two schemes, which we leave to
future work. A future study of our results for more inclusive
samples of $W$ production in association with $b$ jets will also be
interesting (in this study we focus in signatures with exactly 2 $b$ jets).

%%%%%%%%%%%%%%%%%%%%%%%%%%%%%%%%%%%%%%%%%%%%

\section{Results for \Wbb{} production in association with light jets}
\label{sec:vjets}
In this section we present NLO QCD results for \Wbbn~production in
$pp$ collisions at $\sqrt{s}=13$ TeV, the experimental configuration
of the LHC Run-II. We present results for a set of distributions and
apply the following cuts
\begin{align}
  p_T^{\text{jet}}&>25\text{ GeV},& |\eta^{\text{jet}}|&<2.4\ ,\notag\\
  p_T^{e}&>25\text{ GeV},& |\eta^{e}|&<2.5\ ,\notag\\
  p_T^{\nu}&>20\text{ GeV},& M_T^W &> 20\text{ GeV}\ .
  \label{eq:Cuts}
\end{align}
The cuts are applied to both light  and
$b$ jets. The renormalization and factorization scales are chosen to be equal and set on an
event-by-event basis by $\mu=\HTpartonicp/2$, according to
\eqn{eq:htpart}. We define our jets by employing the
anti-$k_T$ jet algorithm~\cite{antikT} with $R=0.4$, as implemented in the
\texttt{FastJet} package~\cite{Cacciari:2011ma}.

\subsection{Total cross sections}
\label{totalxsw}

In \tab{tab_Wpj_total_xs}, we present total partonic cross sections,
employing the kinematical cuts of \eqn{eq:Cuts}, for inclusive production of
both $W^-$ and $W^+$ accompanied by two $b$ jets and zero to three
light jets. The numerical integration uncertainty is given in parenthesis and
the scale dependence is quoted in superscripts and subscripts. We also show
the ratio of NLO over LO results, so called $K$-factors, in separate columns.

%%%%%%%%%%%%%%%%%%%%%%%%%%%%%%%%%%%%%%%%%%%%%% 
\begin{table}[ht]
  \small
  \begin{center}
    \begin{ruledtabular}
      \begin{tabular}{ccccccc}
        jets  & \Wbbm~LO & \Wbbm~NLO & $K$-factor & \Wbbp~LO & \Wbbp~NLO & $K$-factor\\
        \hline
       0  & $0.33278(12)^{+0.0619}_{-0.0490}$ & $0.67719(60)^{+0.1288}_{-0.1000}$  & $2.03$ & $0.48573(19)^{+0.0925}_{-0.0727}$ & $0.97175(85)^{+0.1877}_{-0.1411}$  & $2.00$\\
        1  & $0.36153(13)^{+0.1408}_{-0.0945}$ & $0.50484(63)^{+0.0851}_{-0.0800}$  & $1.40$ & $0.52095(23)^{+0.2034}_{-0.1362}$ & $0.72740(99)^{+0.1277}_{-0.1167}$  & $1.40$\\
        2 & $0.18501(44)^{+0.1053}_{-0.0626}$ & $0.22604(87)^{+0.0407}_{-0.0400}$  & $1.22$ & $0.27663(68)^{+0.1569}_{-0.0934}$ & $0.3340(17)^{+0.0599}_{-0.0647}$  & $1.21$\\
        3  & $0.07204(25)^{+0.0540}_{-0.0289}$ & $0.08288(89)^{+0.0189}_{-0.0200}$  & $1.15$ & $0.11493(59)^{+0.0855}_{-0.0459}$ & $0.1286(17)^{+0.0280}_{-0.0307}$  & $1.12$\\
\end{tabular}
\end{ruledtabular}

     %%%%%%%%%%% TABLE xs  %%%%%%%%%%%%%%%%%%%%%%%%%%
\end{center}
\caption{LO and NLO QCD results for inclusive \Wbbpm+$0,1,2,3$-jet cross
sections (in $pb$). Results with dynamical scale $\HTpartonicp/2$ are shown
together with their respective $K$-factors.  The setup employed is specified in
section~\ref{sec:kin}, and kinematical cuts in \eqn{eq:Cuts}. Scale
dependence is shown in superscripts and subscripts. The number in parenthesis next to
the central value gives the corresponding statistical integration
error.\label{tab_Wpj_total_xs} }
\end{table}

%%%%%%%%%%% TABLE ratios  %%%%%%%%%%%%%%%%%%%%%%%%%%
\begin{table}[ht]
\small
\begin{center}
\begin{ruledtabular}
\begin{tabular}{ccccccc}
\multicolumn{1}{c}{ } & \multicolumn{2}{c}{\Wbbp~$n$/\Wbbm~$n$} &
\multicolumn{2}{c}{\Wbbm~$n/(n-1)$}  &
\multicolumn{2}{c}{\Wbbp~$n/(n-1)$} \\
        \noalign{\vskip 2mm}
        \colrule
        \noalign{\vskip 2mm}
$\qquad n\qquad$ & LO & NLO & LO & NLO  & LO & NLO  \\
        \noalign{\vskip 2mm}
        \colrule
        \noalign{\vskip 2mm}
0 &  $1.45962(78)$ & $1.4350(18)$ & --- &  --- & ---  & --- \\
1&  $1.44098(83)$ & $1.4409(27)$ & $1.08640(55)$ &$ 0.7455(17)$ &$ 1.07253(64)$ & $ 0.7485(12)$ \\
2&  $1.4952(51)$ & $1.4776(95)$ & $0.5117(12)$ &$ 0.4478(21)$ &$ 0.5310(13)$ & $ 0.4592(24)$ \\
3&  $1.5952(99)$ & $1.551(27)$ & $0.3894(16)$ &$ 0.3667(44)$ &$ 0.4155(24)$ & $ 0.3850(54)$ \\
\end{tabular}
\end{ruledtabular}
\end{center}
\caption{LO and NLO QCD cross section ratios. The second and third columns
  give charge ratios for both LO and NLO cross sections as a function of the number of
  associated light jets $n$. The last four columns give jet
  production ratios for both \Wbbm~as well as \Wbbp~in association with $n$ light
  jets. These ratios are taken for the cross section of a given
  process to that with one less jet. The number in parenthesis gives the corresponding statistical integration error.\label{tab_xs_ratios} }
\end{table}
%%%%%%%%%%% TABLE ratios  %%%%%%%%%%%%%%%%%%%%%%%%%%

LO cross sections display a large scale sensitivity, reaching up to 60\% for
\Wbbjjj{} production. We note that the scale dependence of the LO cross section
for \Wbb{} is around $20\%$ while the NLO QCD corrections increase the
total cross section by a factor of 2. This clearly highlights that scale
dependence is in general not representative of the associated theoretical
uncertainties. In this case, the large quantum corrections can be understood as a
result of the opening of gluon-initiated
channels~\cite{Ellis:1998fv,FebresCordero:2006sj,Cordero:2009kv}. Also for \Wbbj{} a gluon-gluon initiated channel is opened
up, but with milder impact, and for the larger multiplicity processes all
subprocesses are present at LO. Hence, quantum corrections are milder
for these processes. Furthermore, kinematical
constraints at LO are only present for \Wbb{} production, as we will discuss for example for the $p_T^{b\bar b}$ and
$p_T^W$ observables in section~\ref{sec:hw}. As a consequence, we expect quantum corrections for
processes with even more light jets to be under relatively good pertubative control.

In \tab{tab_xs_ratios} we show first, in columns 2 and 3, $W^+/W^-$ charge
ratios as a function of the number of jets. These ratios show a large stability
with respect to the quantum corrections, which have been explored in similar
processes as a way to make precise determinations of ratios of $u/d$ PDFs (see
for example ref.~\cite{Kom:2010mv}). They also show some stability as a function
of the number of jets, with a slight monotonic increase given the larger mean
values of Bjorken $x$ sampled as a consequence of the larger invariant mass necessary to
produce the corresponding final states.

Finally we also explore in \tab{tab_xs_ratios} the jet ratios in $W^\pm
b\bar b$ production in association with light jets. Similarly to studies of these ratios in $W+n$-jet (light jet)
production~\cite{BH:Wratios}, we observe that the results for $n=1$ are special
given the large NLO corrections for \Wbb{} production. The opening of an
initial-state channel makes the
\Wbbj{}$/$\Wbb{} ratio clearly sensible to quantum corrections. In the light jet
study~\cite{BH:Wratios} this was the case for the ratio ($W+2$-jet$)/$($W+1-$jet),
and a full study of jet-ratio universality needed the completion of the NLO QCD
correction to $W+5$-jet production~\cite{BH:W5j}. Similarly, in \Wbb{} inclusive
production, it might be interesting to explore the NLO QCD corrections to
$W+b\bar b+4$-jet production in the future.

\subsection{Scale Dependence}\label{wscale}
\begin{figure}[tbh]
\begin{center}
\includegraphics[clip,scale=0.71]{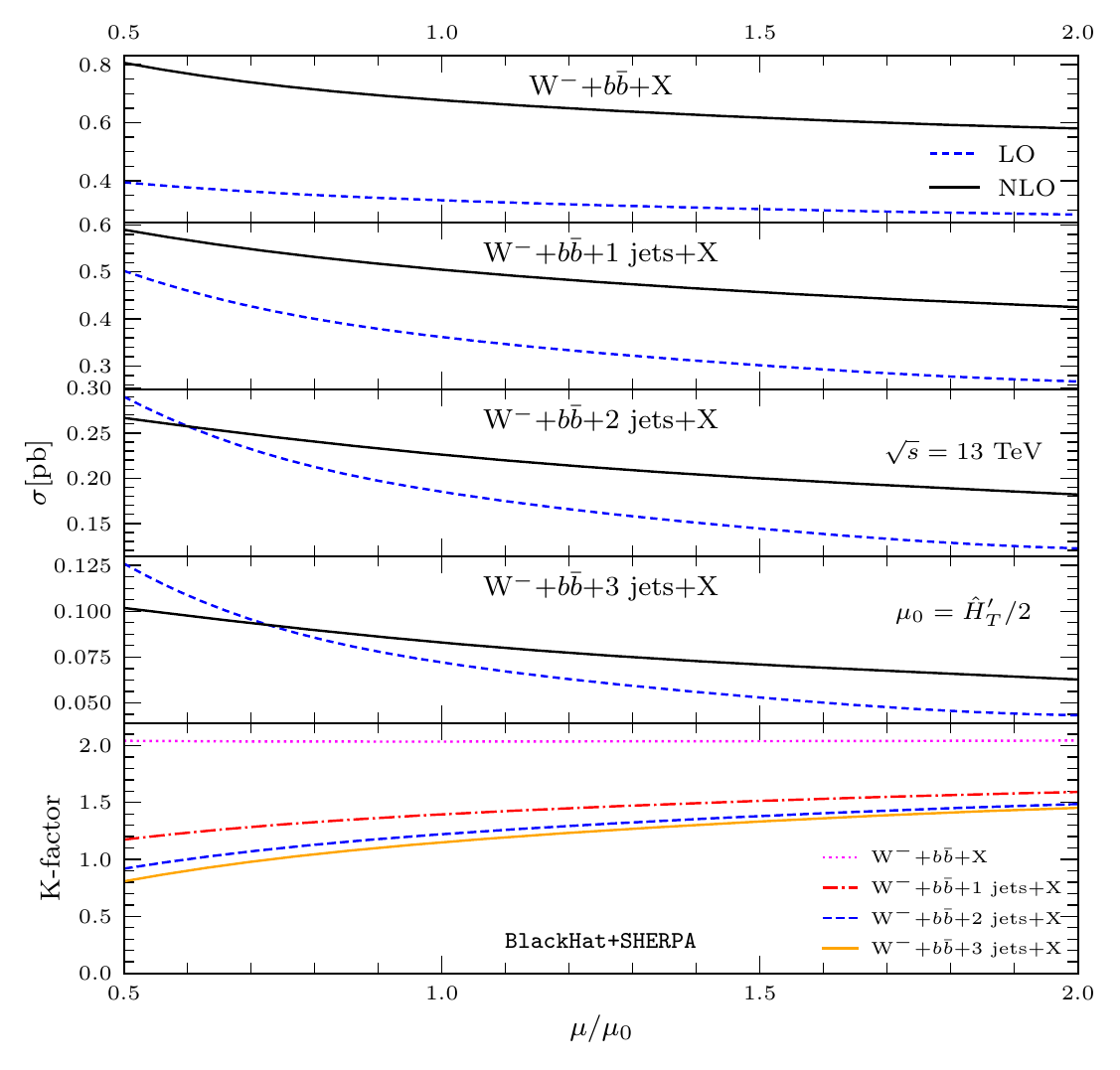}
\includegraphics[clip,scale=0.71]{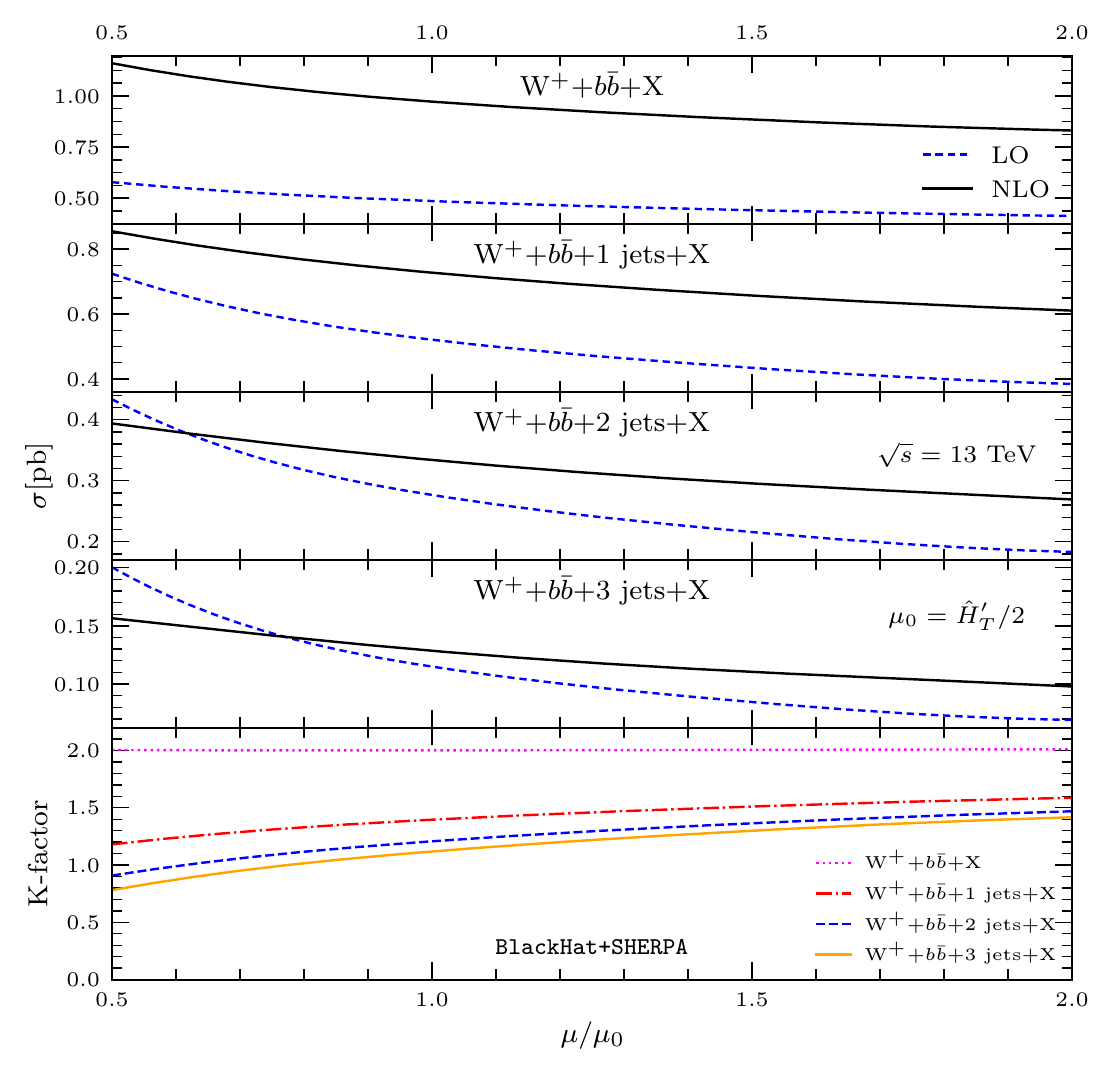}
\end{center}
\caption{The renormalization- and factorization-scale dependence of total cross
  sections for \Wbbm$+0,1,2,3$-jet$+X$ production in the left and
\Wbbp$+0,1,2,3$-jet$+X$ production to the right,
 with $\mu_0=\mu_{\rm r}=\mu_{\rm f}=\HTpartonicp/2$. 
The upper four panels show the dependence of LO (dashed blue line) and
  NLO (solid black line) predictions. The lower panel shows
  the K-factor (ratio of NLO/LO).}
\label{fig_Wjets_sdep}
\end{figure}

In \fig{fig_Wjets_sdep} we study the dependence of total cross sections in
\Wbbm{} and \Wbbp{} production in association with up to 3 light jets on the
renormalization and factorization scale. We employ the central dynamical scale
$\mu_0=\mu_{\rm R}=\mu_{\rm F}=\HTpartonicp/2$. The scale variations
observed for $W^+$ and for $W^-$ are very similar. The LO cross sections have a monotonically increasing
scale dependence, for $n\geq 1$.  As we observed in the previous subsection, the
scale dependence of \Wbb{} production is special.

%scale dependence for Wm
%%%%%%%%%%%%% FIGURE %%%%%%%%%%%%%%%%%%
%%%%%%%%%%%%%%%%%%%%%%%%%%%%%%%%%%%%%%%

We choose the dynamical scale $\HTpartonicp/2$ which on average increases monotonically with multiplicity. For vector boson production in association with massless jets this scale choice
has been observed to produce stable NLO results over a wide range of kinematical
configurations relevant to the LHC and future
colliders~\cite{BH:W3jPRL,BH:W4j,BH:W5j,Mangano:2016jyj}. For the LHC in
particular, it has been observed that for massless jet production the scale
$\HTpartonicp/2$ typically produced NLO cross sections lying on the locus of the
scale-dependence curves. Here we observe that for \Wbbn{} production, the
NLO cross section at the central scale appears consistently on the right of the
scale-dependence plateau. We can assert, in particular considering the
similarities of the massive and massless results studied in
section~\ref{sec:bmass}, that this difference has little to do with the presence
of a massive jet, and it is actually due to the dominant type of subprocess. For
light-jet production those are the ones with a single quark line, while in
the case of \Wbbn{} production the dominant subprocess are those with
two quark lines (those are the subprocesses with most gluons allowed).

%pt leading bjet
%%%%%%%%%%%%% FIGURE %%%%%%%%%%%%%%%%%%
\begin{figure}[ht]
\begin{centering}
\includegraphics[clip,scale=1]{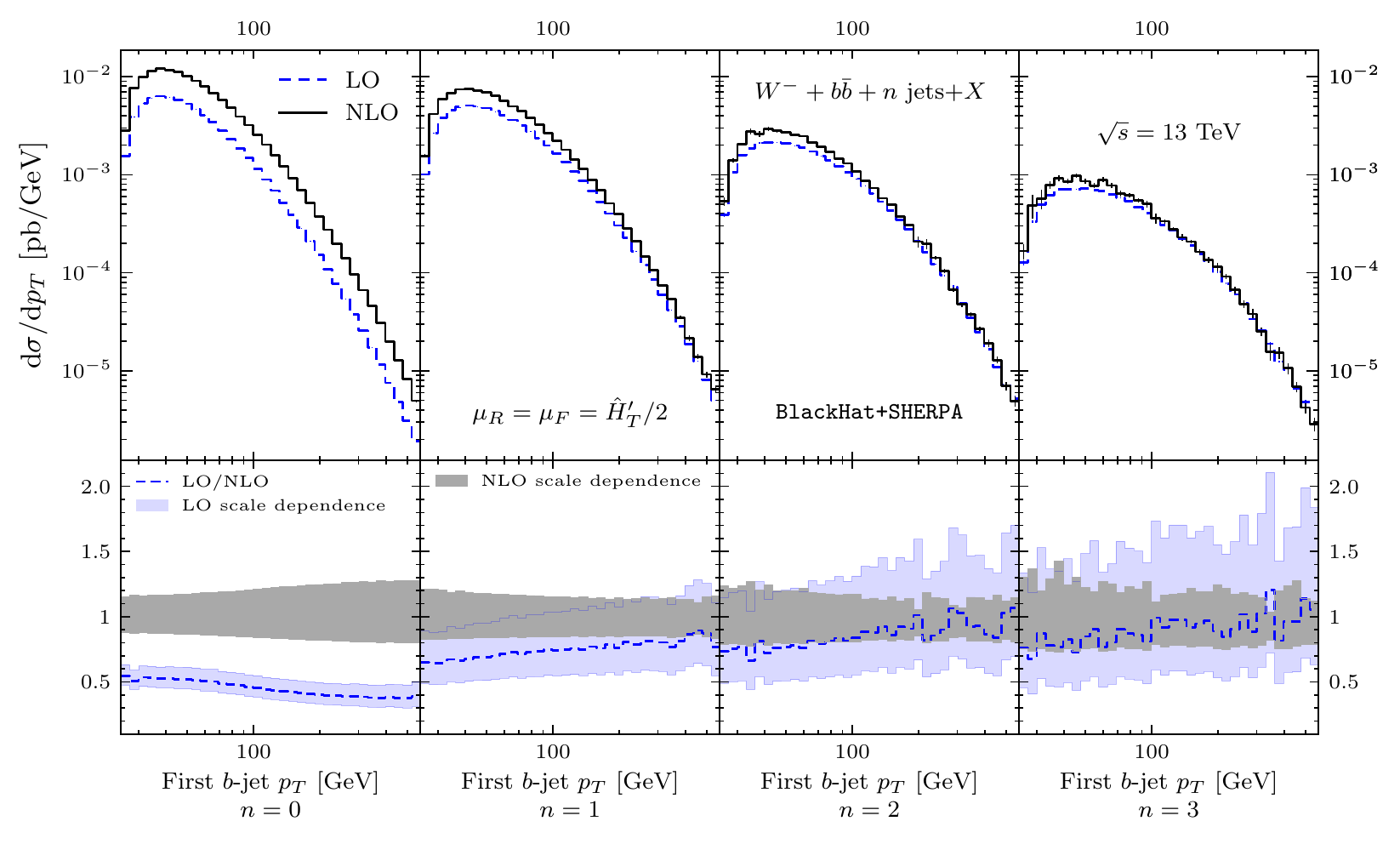}
\end{centering}
  \caption{The $\pT$ distributions of the leading $b$ jet (ordered by $p_T$) in inclusive \Wbbm$+n$-jet
    production at the LHC with $\sqrt{s}=13$~TeV. The light-jet multiplicity 
	increases from $n=0$ to $n=3$ from left to right. In the upper panels the
    dashed (blue) lines show the LO results and the solid (black) lines the NLO
    results. Vertical thin lines show the statistical error from the numerical
    integration. In the bottom panels we show the scale-dependence bands
    normalized to the NLO result, in blue for LO and dark gray for NLO.}
  \label{fig_Wmnjpt}
\end{figure}
%%%%%%%%%%%%%%%%%%%%%%%%%%%%%%%%%%%%%%%

%pt subleading bjet
%%%%%%%%%%%%% FIGURE %%%%%%%%%%%%%%%%%%
\begin{figure}[ht]
\begin{centering}
\includegraphics[clip,scale=1]{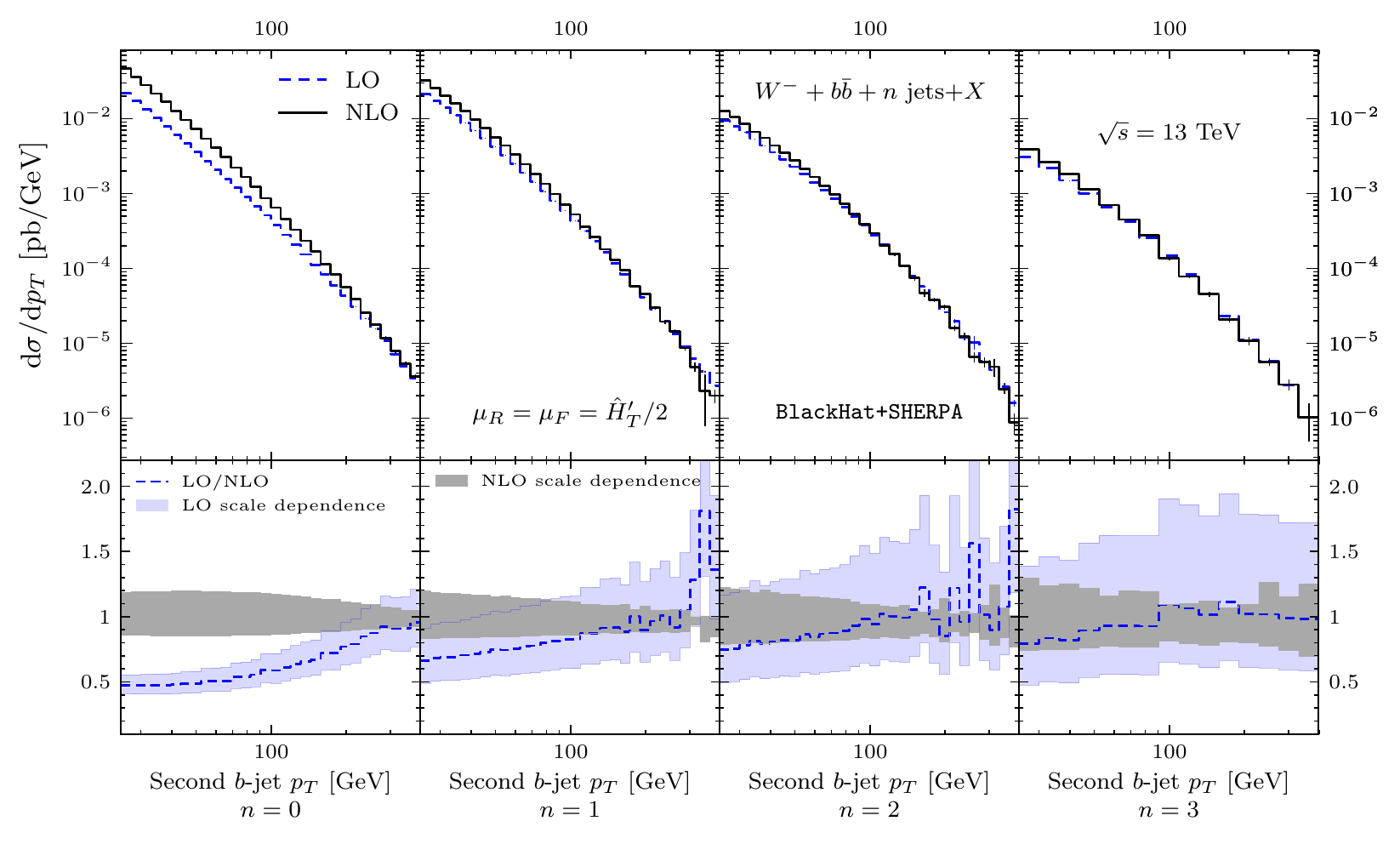}
\end{centering}
  \caption{The $\pT$ distributions of the subleading $b$ jet (ordered by $p_T$) in inclusive \Wbbm$+n$-jet
    production at the LHC with $\sqrt{s}=13$~TeV. Format as in \fig{fig_Wmnjpt}.}
  \label{fig_Wmnjpt2}
\end{figure}
%%%%%%%%%%%%%%%%%%%%%%%%%%%%%%%%%%%%%%%

Another interesting difference between $W$ production in association with light
jets and \Wbb{} production with multiple light jets, is that for the
former the leading-color approximation for one-loop matrix elements gave a very good
approximation for physical observables (at the level of 1 to
3\%). Contrary to that, \Wbb{} production with light jets is largely dominated by virtual
contributions in our setup, and so the leading-color approximation is at the
order of 10\% for physical observables. That is why all of our results in this
article include full-color information, and we only
exploit the decomposition in a color expansion for efficiency of the
computation. We again attribute this difference to the unlike dominant
subprocesses.

\subsection{Differential distributions}
\label{diffxsw}

In this section, we describe NLO results for several differential
distributions and thereby analyze the impact
that quantum corrections have on fixed-order predictions over phase
space. We generally show results only for one of the $W^\pm$ charges, as the
structure of the corrections are similar between them. 

In \figs{fig_Wmnjpt}{fig_Wmnjpt2} we show the jet-$p_T$ spectra of the leading
and subleading $b$ jets (ordered by $p_T$) respectively, for inclusive
\Wbbm{} production in association with $n=0,1,2,3$ jets. The upper panel
of the
figures show the LO and NLO distributions in dashed (blue) and solid (black)
lines respectively, while the bottom panels show the scale-dependence bands
normalized to the central NLO result (LO in blue and NLO in gray). Numerical
integration errors for each bin are shown as thin vertical lines (when visible).
All distributions will be shown in a similar manner.

The NLO corrections show quite some structure beyond the $K$-factors studied at
the level of the total cross sections in the previous subsection. We
observe shape differences in most of the $p_T$ distributions of the $b$ jets, in a way
that make the LO predictions usually harder (with the exception of the leading
$b$ jet $p_T$ in \Wbb{} production). Nevertheless, the LO/NLO shape difference
is clearly reduced for the process with highest multiplicity, \Wbbjjj{} production, a
feature that shows up persistently in the following observables.
We notice that the scale dependence of the NLO
results is reduced compared to the LO results (apart from \Wbb{}, as
discussed for total cross sections). In the high multiplicity samples, the NLO results
lie inside the LO bands. 

%Softest light-jet PT
%%%%%%%%%%%%% FIGURE %%%%%%%%%%%%%%%%%%
\begin{figure}[ht]
\begin{centering}
\includegraphics[clip,scale=1]{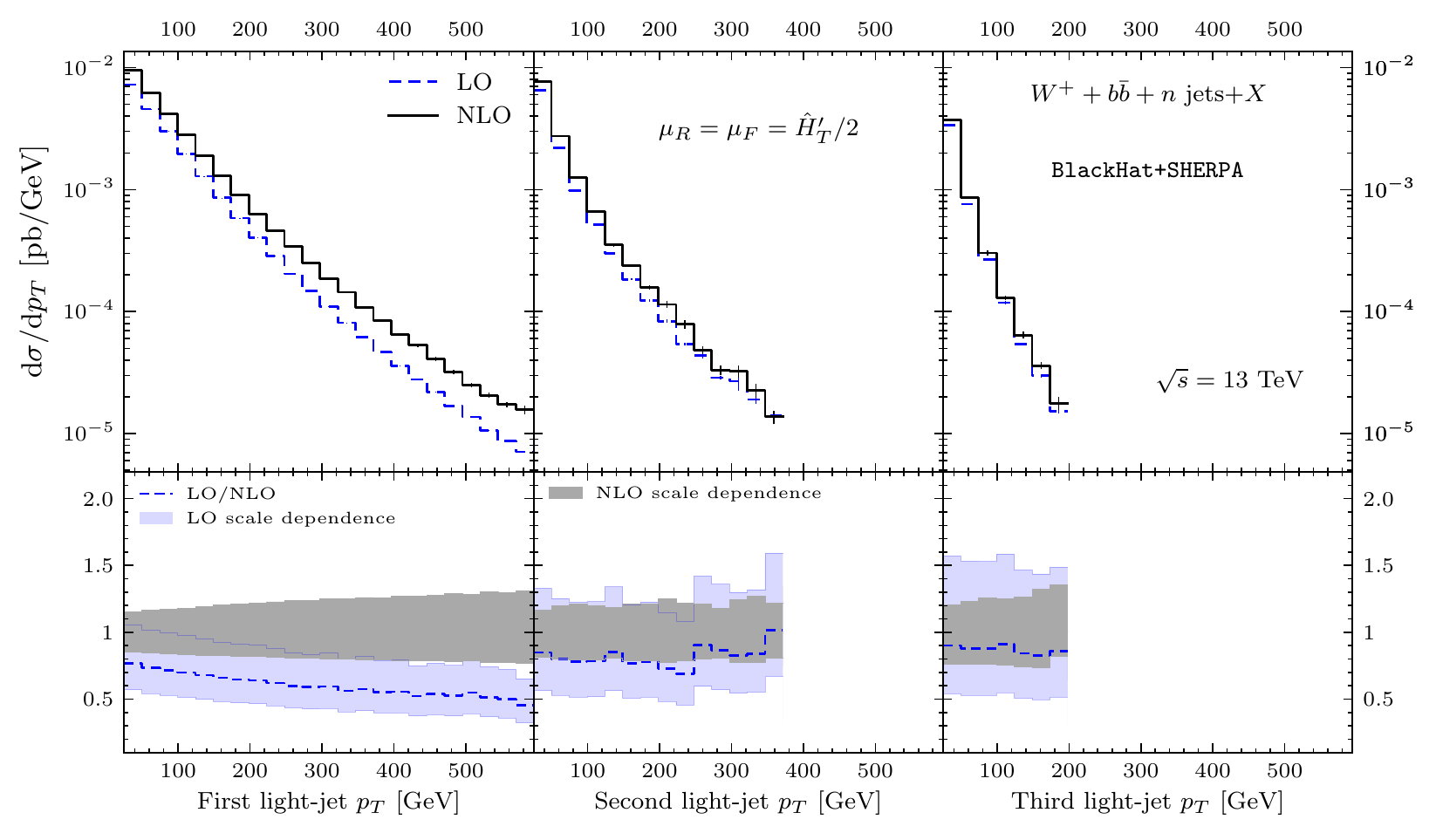}
\end{centering}
  \caption{The $\pT$
distributions of the softest light jet in inclusive \Wbbp$+n$-jet production.
Format as in \fig{fig_Wmnjpt}.}
  \label{fig_Wmnjptlight}
\end{figure}
%%%%%%%%%%%%%%%%%%%%%%%%%%%%%%%%%%%%%%%

In \fig{fig_Wmnjptlight} we show the $p_T$ distributions of the softest light jet in
inclusive \Wbbp{}$+1,2,3$-jet production. We observe a considerable reduction of the scale
sensitivity with the inclusion of the QCD corrections, with overlap of the LO
and NLO bands. It is important to note that for these distributions, which are
experimentally very relevant as they are quite sensitive to the jet-energy
scale, the quantum corrections are rather flat. The feature is similar to
what has been observed for softest jet $p_T$ distributions in $W+n$-light-jet
production, and which is associated to the choice of renormalization and
factorization scales $\HTpartonicp/2$ in the LO result. Notice that the NLO
results are rather insensitive to the choice of dynamical scale, as long as the
choice is naturally connected to the kinematic configurations of the processes
under study.

%HT jet / hadronic
%%%%%%%%%%%%% FIGURE %%%%%%%%%%%%%%%%%%
\begin{figure}[ht]
\begin{centering}
\includegraphics[clip,scale=1.0]{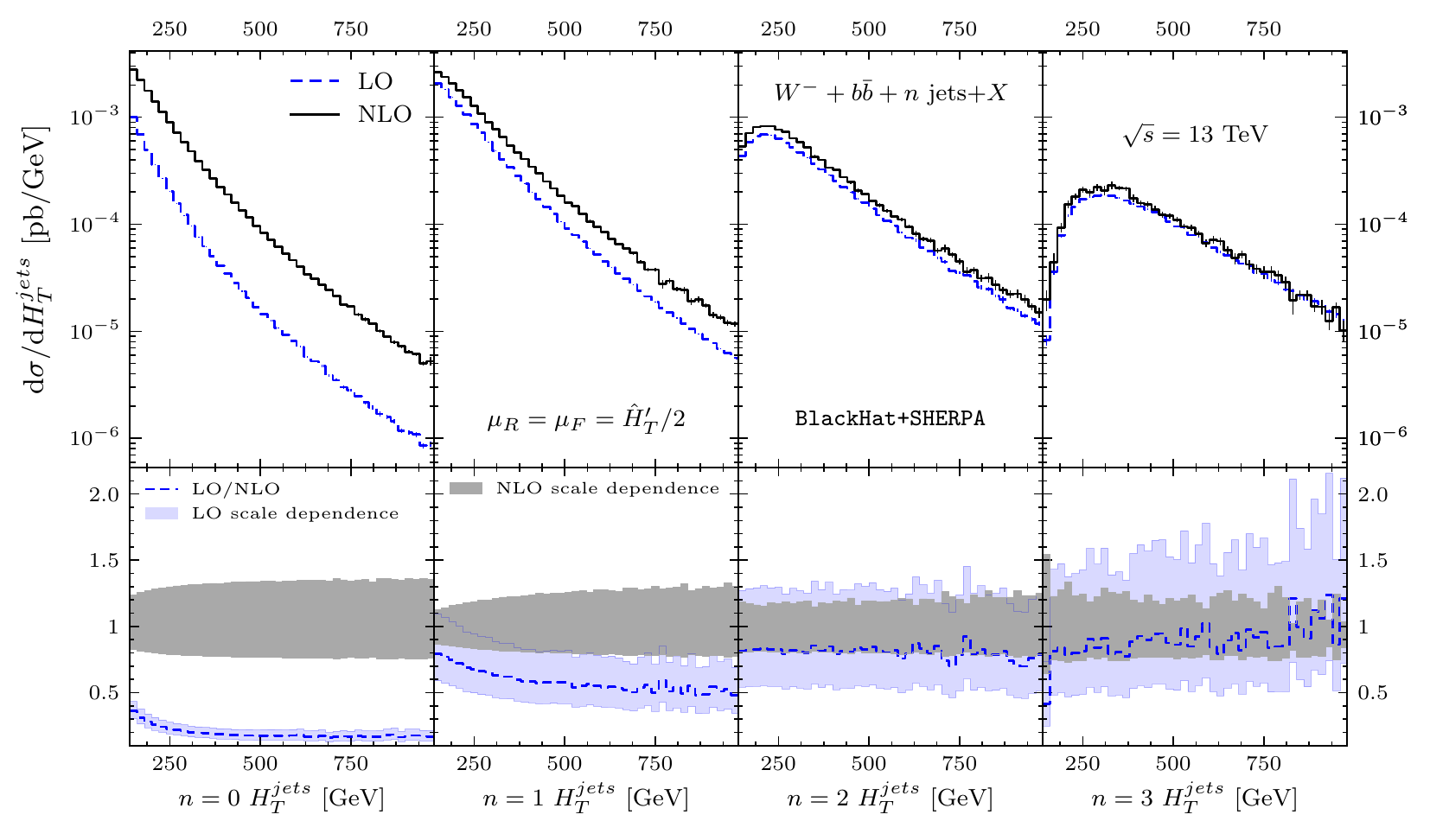}
\end{centering}
  \caption{Distribution in the total transverse jet energy
    $H_T^{jets}$ of light  and $b$ jets for inclusive \Wbbm$+n$-jet
    production at the LHC with $\sqrt{s}=13$~TeV. Format as in \fig{fig_Wmnjpt}.}
  \label{fig_Wmnjht}
\end{figure}
%%%%%%%%%%%%%%%%%%%%%%%%%%%%%%%%%%%%%%%

An interesting observable for many scenarios of physics beyond the SM (BSM), as
well as for experimental studies at hadron colliders, is that of the total
hadronic activity in a detector. In \fig{fig_Wmnjht} we show the distribution in
this observable, including all hadronic activity from the light and $b$ jets in
\Wbbm$+0,1,2,3$-jet production. Large and phase-space dependent NLO
corrections appear for \Wbb{} as we would expect from previous discussions.
Interestingly, in \Wbbj{} production a remnant of these large effects appears
in this observable. The corrections are not as large as for the former, but still at
around 1~TeV for $H_T^{jets}$, we see a differential $K$-factor reaching two,
though the shape difference seems to end at about 400~GeV. The
large-multiplicity processes on the other hand show much less
structure, related to the kinematically unconstrained nature of their
LO configurations, which contain any $W$ soft enhancements starting at LO.

%dR first b-jet / charged lepton
%%%%%%%%%%%%% FIGURE %%%%%%%%%%%%%%%%%%
\begin{figure}[ht]
\begin{centering}
\includegraphics[clip,scale=1.0]{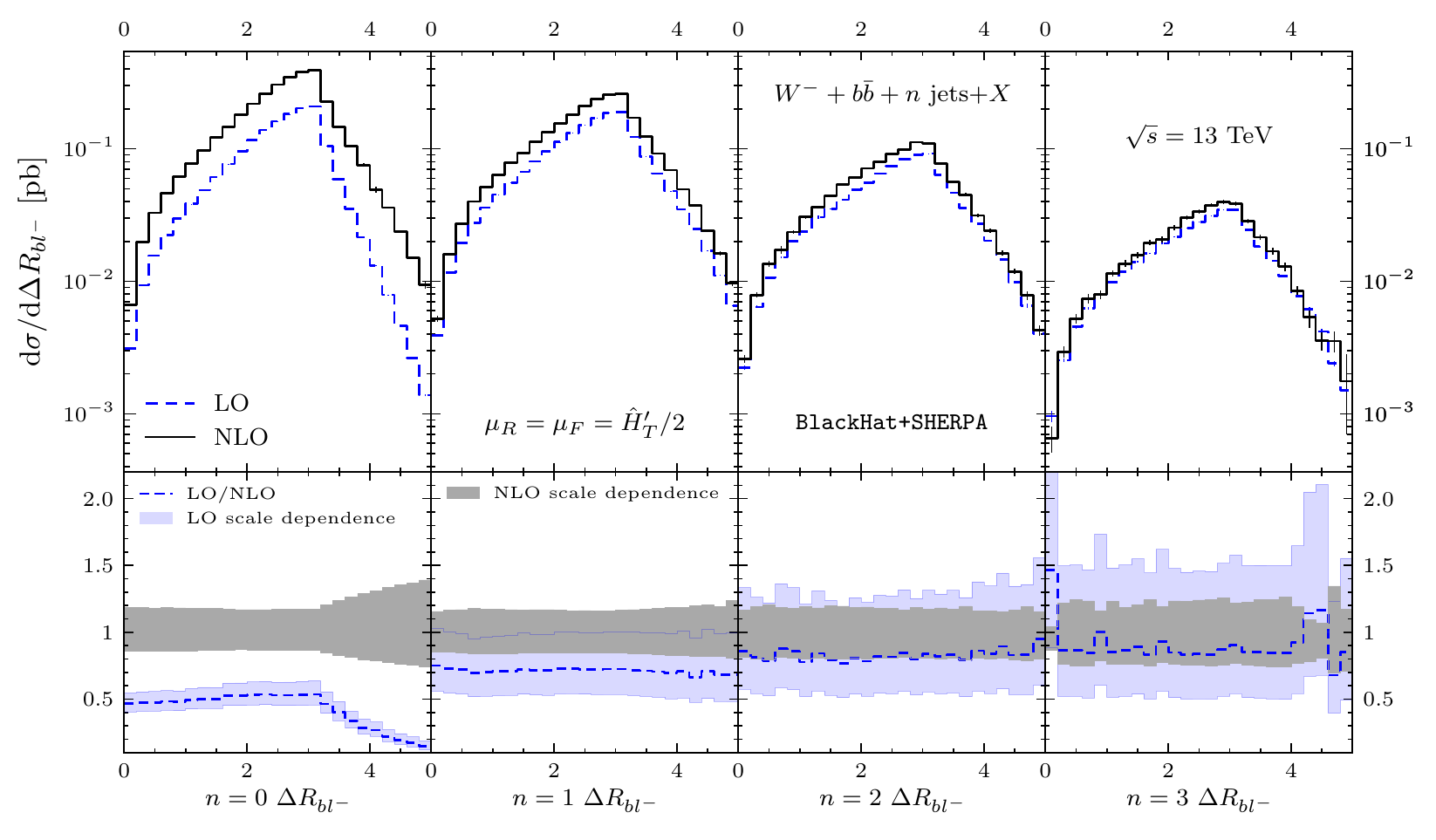}
\end{centering}
  \caption{Distribution in the $\Delta R_{bl^-}$ separation between the first
    $b$ jet (ordered in $p_T$) and the charged lepton for inclusive \Wbbm$+n$ jets
    production.  Format as in \fig{fig_Wmnjpt}.}
  \label{fig_Wmnjdrbl}
\end{figure}
%%%%%%%%%%%%%%%%%%%%%%%%%%%%%%%%%%%%%%%

Finally, to end this section, we show in \fig{fig_Wmnjdrbl} the distribution
on the $\Delta R$ separation between the first $b$ jet and charged lepton $l^-$. Most
of the angular variables that we have studied are similar to this one, which
shows little structure in the QCD corrections. We only find effects
when a certain kinematic constraint is imposed at LO and released by the corrections,
as it is the case on the left most plots of \fig{fig_Wmnjdrbl}. 
In the case of the $\Delta R_{bl^-}$ at LO in \Wbbm{} production, the parent $W$
and gluon that give rise to the leptons and $b$ jets are produced with $\Delta
\phi$ (the difference in azimuthal angle) equal to $\pi$. Also, the $\Delta \eta$
distribution peaks at around zero and decreases monotonically. The resulting $\Delta R_{bl^-}$ distribution thus has the feature of a sharply decaying
distribution at LO. All
those constrains are lifted by the real corrections and do not appear at all in
\Wbb$+1,2,3$-jet production.

\section{Backgrounds to $HW$ production}
\label{sec:hw}

So far all LHC measurements of Higgs-boson properties appear in good agreement
with predictions of the SM 
(see for
example ref.~\cite{Khachatryan:2016vau}). One of
the properties that will be important to constrain further is the coupling strength of the Higgs
boson to $b$ quarks. Given that a Higgs boson with mass $M_h$ around 125 GeV is
supposed to decay more than half of the time into a $b\bar b$ pair, it is of great
importance to constrain the Yukawa coupling $y_b$ and consequently 
learn about the Higgs boson's total decay width.
In the main
production channel of the Higgs, through gluon-gluon fusion, one faces
a large background from pure QCD $b\bar b$ production. Therefore,
considering the associated $Wb\bar b$ production gives an extra handle to
detect the Higgs decaying to $b$ quarks (see the recent measurement by ATLAS in
ref.~\cite{ATLAS:hbb2017}). This of course, as long as the
irreducible backgrounds for \Wbb{} production in the SM can be kept under
control. The predictions provided in this article aim to contribute
to these studies.

Some of the key observables for $HW$ analyses are those associated to the $b\bar
b$ system, in particular $p_T^{b\bar b}$ and $M_{b\bar b}$. When
producing a Higgs, those are associated with the $p_T$ distribution
and the invariant mass of the Higgs. In addition,
distributions that help to characterize the accompanying $W$ boson are important, for
example $p_T^W$. In this section we study those three observables 
with our high-multiplicity results.
 At high energies the presented spectra are
enhanced with resonant top contributions (see for example the recent
study~\cite{Denner:2017kzu}). Nevertheless, in the context of $HW$ production
the focus is on the non-resonant backgrounds. The non-resonant top contributions
are sizable, and can be of similar order to the ones presented here. We leave
studies of non-resonant top contributions to future work.

%pt(bb) 
%%%%%%%%%%%%% FIGURE %%%%%%%%%%%%%%%%%%
\begin{figure}[ht]
\begin{centering}
\includegraphics[clip,scale=1]{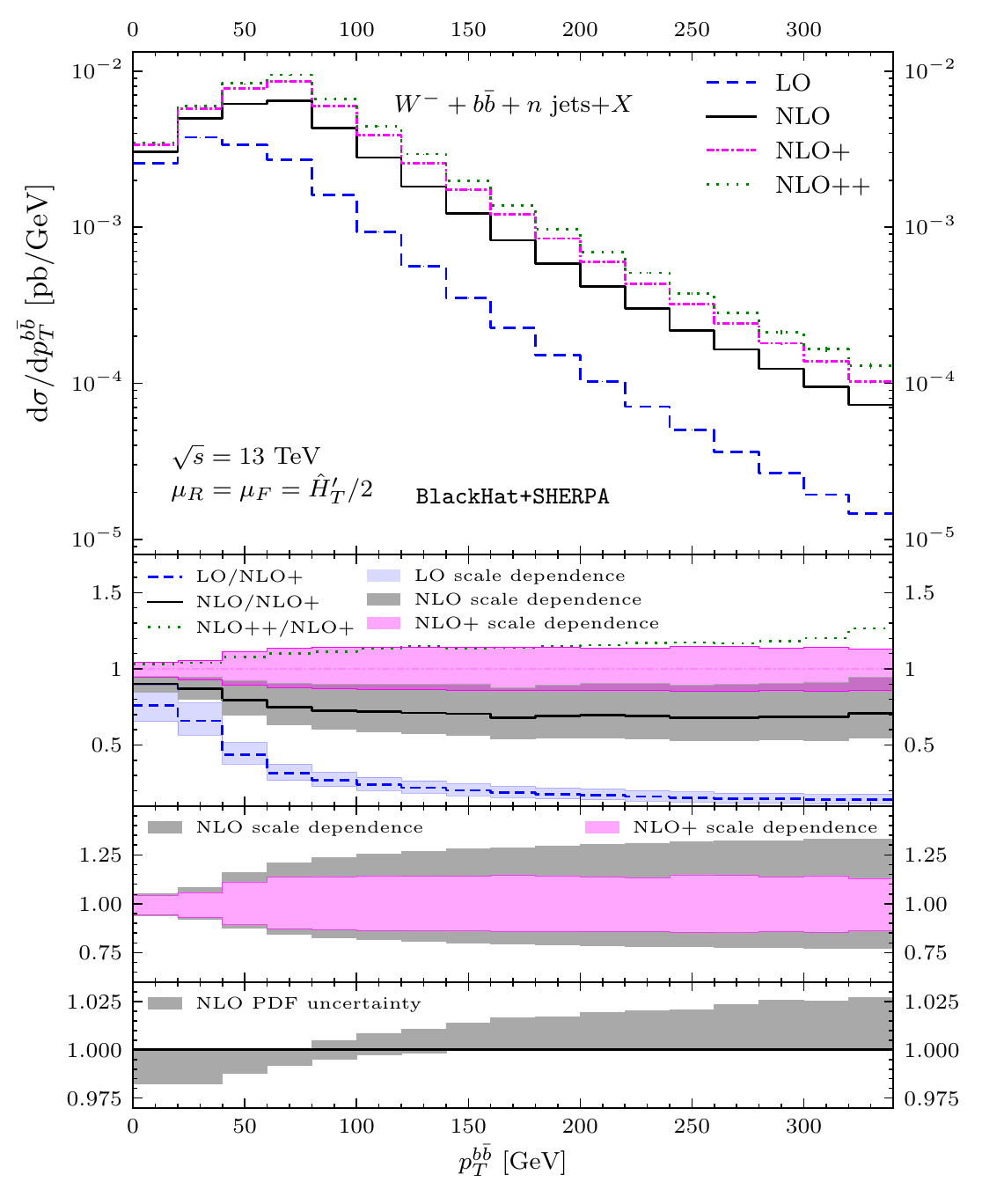}
\end{centering}
  \caption{The $p_T$ distribution of the $b\bar{b}$ system in inclusive \Wbbm{} production,
    computed at LO (dashed blue line) and NLO (solid black line) as well as
    by employing the exclusive sums NLO+ (dashed-dot magenta line)
    and NLO++ (dotted green line).
    The second panel shows scale dependence bands normalized by NLO+,
    and in the third panel they are normalized by the corresponding
    central value. The bottom panel shows the associated PDF uncertainties
    normalized to our NLO results.}
  \label{fig_Wmnjptbb}
\end{figure}
%%%%%%%%%%%%%%%%%%%%%%%%%%%%%%%%%%%%%%%

The NLO QCD correction to \Wbb{} production have large contributions associated to
processes with an extra light jet~\cite{Ellis:1998fv,FebresCordero:2006sj,Cordero:2009kv}.
 A way to handle those
contributions was to obtain exclusive results in the number of light
jets~\cite{FebresCordero:2006sj}, which explicitly vetoed
events with extra jets, but that prescription suffers from its sensitivity
to the $p_T^{\rm veto}$ cut~\cite{Tackmann:2012bt}. 
In this article we consider a different approach, using the `exclusive sums'
technique~\cite{ESums}.
Instead of imposing a veto cut to stabilize the predictions, this approach
replaces the extra light-jet contributions to generic observables,
which are effectively LO, by their corresponding results including NLO QCD
corrections. In higher-order corrections such contributions are naturally
added. However, as they are hard to obtain, we use the above
approximation and analyze the impact in our predictions.

The `exclusive sums' technique is expected to give improved predictions when
tree-like contribution, with an extra light jet, to NLO corrections are 
large. Notice that in measurements of $W$+light jets some of the predictions
from exclusive sums have been compared to LHC data, see for
example~\cite{Aad:2014qxa,ATLAS:ratio2017}, usually in the context of $W+1$-jet production. By
now those computations are outdated, given the recent NNLO QCD calculation
presented in ref.~\cite{Boughezal:2015dva}. It is important to mention that for the comparison to LHC data the
application of a parton-shower can be studied \cite{Luisoni:2015mpa}, or using a
matched and merged version for example with the MEPS@NLO \cite{Hoeche:2012yf} or FxFx technique \cite{Frederix:2012ps}.

We will focus on predictions for \Wbb$+X$ production. Fixed-order
results for those will be denoted as usual with the labels LO and NLO. The
exclusive sums we employ are defined in Eq.~\eqref{eq:excsums}, for which we
will use the labels NLO+ and NLO++.
We will use the latter only as a proxy for the size of even
higher-order corrections, that is as an estimator of theoretical uncertainty.
Our main predictions will be that of NLO+.

For completeness, to characterize theoretical uncertainties, we will also explore the PDF uncertainty associated to the
observables under consideration, even though 
they turn out to be subleading. In order to estimate the PDF uncertainty we use
the error sets from the pseudo-PDF set
\texttt{PDF4LHC15\_nlo\_nf4\_30}~\cite{Butterworth:2015oua}. 
Given the smallness of the PDF errors compared to
other theoretical uncertainties, we do not go beyond this
restricted error set for our estimations. Other sources of
uncertainties are the values of $m_b$ and $\alps$, but those are expected to
be rather small, e.g. when compared to missing higher-order corrections.

%pt(W) 
%%%%%%%%%%%%% FIGURE %%%%%%%%%%%%%%%%%%
\begin{figure}[ht]
\begin{centering}
\includegraphics[clip,scale=1]{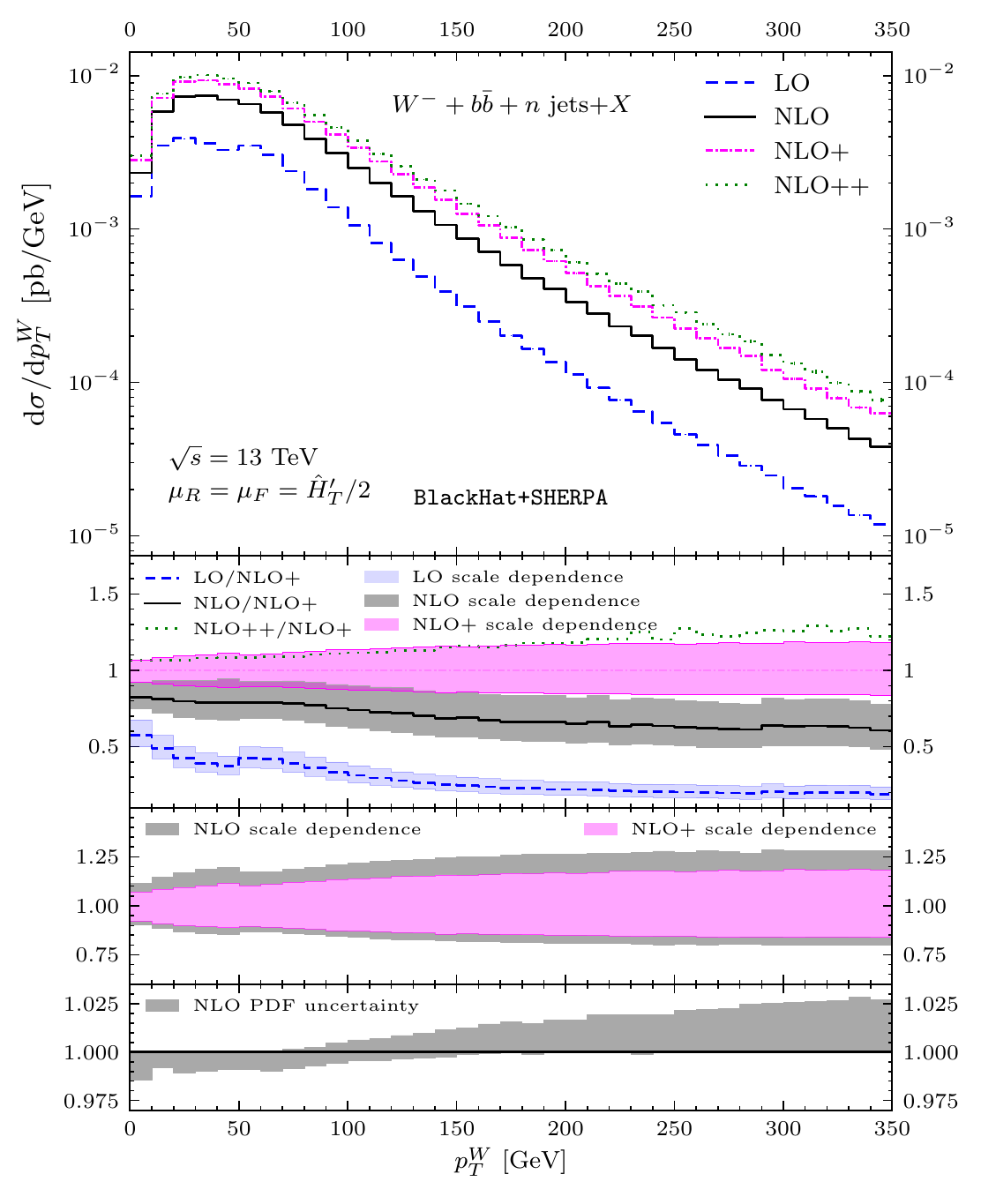}
\end{centering}
  \caption{The $p_T$ of the $W$ boson in inclusive \Wbbm{} production. Format as in \fig{fig_Wmnjptbb}.}
  \label{fig_Wmnjptw}
\end{figure}
%%%%%%%%%%%%%%%%%%%%%%%%%%%%%%%%%%%%%%%

In \fig{fig_Wmnjptbb}, we show the transverse momentum distribution of
the $b\bar b$ system. In the upper panel, we show all of our predictions, including the central `NLO+' prediction as well as LO, NLO
and NLO++. The second panel shows the corresponding scale-dependence bands at
LO, NLO and NLO+, all normalized by NLO+, as well as the central value for
NLO++. In the third panel we show the scale-dependence bands at
NLO and NLO+, normalized by their corresponding central value. In the bottom panel we show the PDF uncertainties, which are always below
2\% for all the ranges of $p_T^{b\bar b}$ shown (we normalized the PDF
uncertainties by our central NLO result). The scale-dependence bands
of the NLO+ predictions are at the level of 13\%, which is a reduction compared to the 26\%
of the fixed-order NLO result. The LO result gives no adequate
prediction. The NLO and NLO+ bands overlap, though they show a difference in
shape, particularly at low values of $p_T^{b\bar b}$. We find that the
higher-order corrections estimated through scale variations and by the NLO++
results are of the same order of magnitude, at the level of 10\%.

For the $p_T^{b\bar b}$ observable, one obseves the release of
kinematical constraints at NLO (which tights up $p_T^{b\bar b}$ and $p_T^W$).
Since at LO the massive $b$-quark pair
originates in a gluon splitting, the kinematical constraints in \Wbb{} are
similar to those appearing in $V+1$ light jet (see e.g.
refs.~\cite{BH:W3jPRL,BH:W4j,BH:W5j}). Real-radiation emission relaxes this
constraint and yields large corrections at NLO through a soft enhancement, which
gives rise to a giant $K$-factor~\cite{Rubin:2010xp}. These characteristics of
the NLO results for \Wbb{} production are then expected~\cite{Catani:1997xc} and
require resummation although fixed order corrections are known to improve the
predictions~\cite{Ridder:2015dxa}.

%Mbb 
%%%%%%%%%%%%% FIGURE %%%%%%%%%%%%%%%%%%
\begin{figure}[ht]
\begin{centering}
\includegraphics[clip,scale=1]{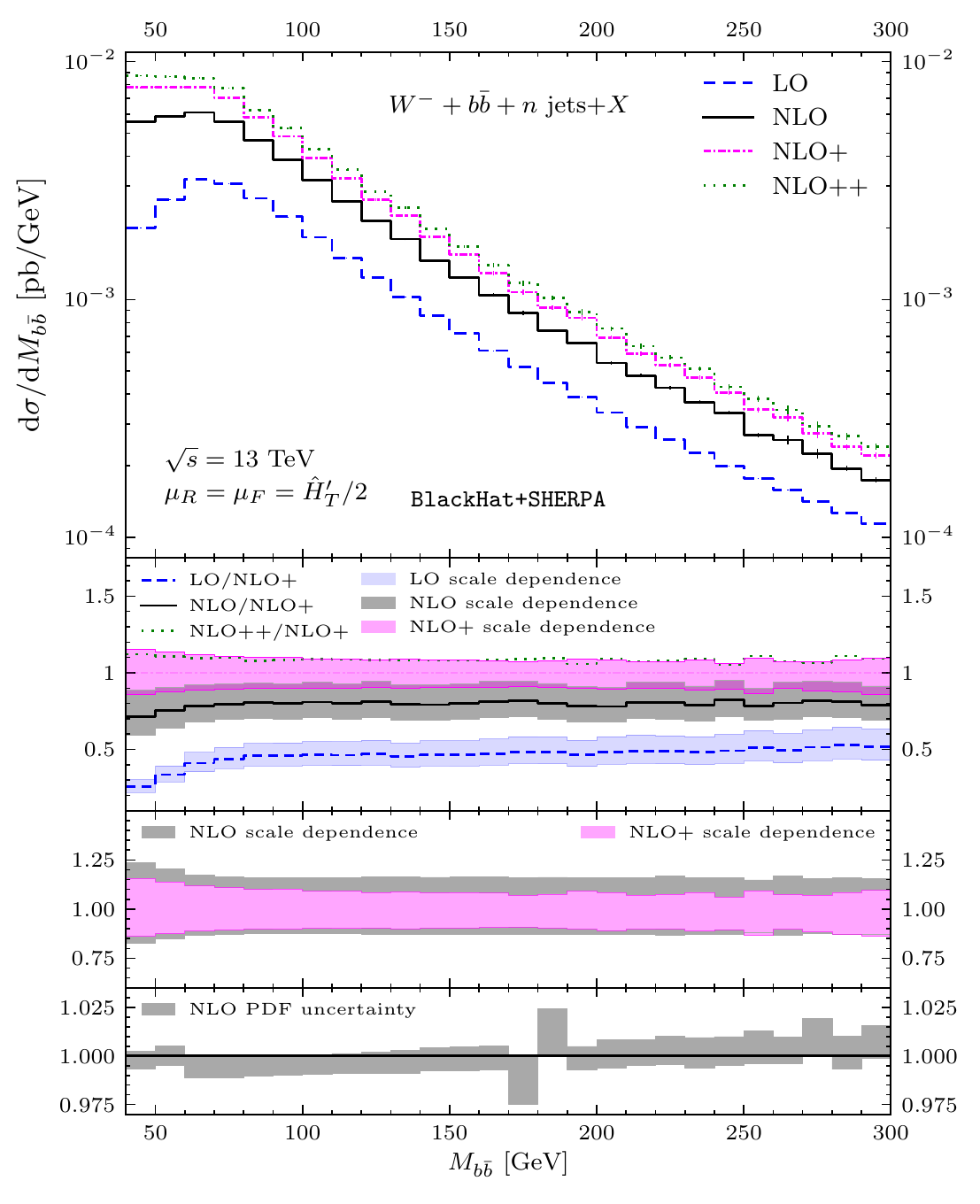}
\end{centering}
  \caption{The invariant mass of the $b\bar b$ system in inclusive \Wbbm{} production. Format as in \fig{fig_Wmnjptbb}.}
  \label{fig_Wmnjmbb}
\end{figure}
%%%%%%%%%%%%%%%%%%%%%%%%%%%%%%%%%%%%%%%

In \fig{fig_Wmnjptw}, we show in the same manner the distribution in the transverse
momentum of the vector boson $p_T^W$. The results are again similar to what we
encounter for $p_T^{b\bar b}$, with the NLO+ uncertainty estimation marginally
overlapping with the NLO predictions and with its scale sensitivity of the order
of the NLO++ predictions. The estimation of theoretical uncertainties is about 17\% in the
range of $p_T^W$ shown (as compared to the 25\% of the NLO results). Again, the PDF uncertainties are
subleading, appearing at 3\% and below.

Finally in \fig{fig_Wmnjmbb} we show the distributions in the $M_{b\bar b}$
observable, which exhibit similar features to the observables
studied previously in this section. In this case, the uncertainties associated with scale
sensitivity and missing higher-order effects appear at about 10\% or slightly
smaller. The PDF uncertainties are tiny, being at 1\% or smaller.

\section{Conclusions}
\label{sec:conclusions}
In this paper we presented the first NLO QCD results for \Wbbjj{} and \Wbbjjj{}
production. For completeness we have also included results with zero or
one extra light jet at NLO QCD. These are the first results obtained with a
new version~\cite{BlackHatII} of the \BlackHat{} library \cite{BlackHatI} which has been constructed to handle
massive particles in one-loop matrix elements. 

We have found that NLO QCD corrections for \Wbbjj{} and \Wbbjjj{} production are mild, with the associated K-factors dropping from 1.4 for \Wbbj{}, to 1.2 for \Wbbjj{}, and 1.15 for \Wbbjjj{} production.
For the latter process the remaining
renormalization and factorization scale 
dependence amounts to 60\% at LO and 20\% at NLO for 
the typical variation in a correlated or uncorrelated way of the scales by a factor of two.
In contrast to this the  K-factor for NLO QCD corrections is large in inclusive
\Wbb{} production~\cite{Ellis:1998fv,FebresCordero:2006sj,Cordero:2009kv}. 
This difference led us to explore possible
improvements to observables for the \Wbb{} process by using exclusive sums, in particular in the context of
$H(\rightarrow b\bar b)W$ associated production.
We
showed that exclusive-sum predictions for key observables give uncertainties
associated to missing higher-order corrections at the level of 10\% to 17\%.
We
found that uncertainties associated to PDFs are subleading, at the level of 2\%
in the generic kinematical regimes studied. 

The ATLAS and CMS collaborations have performed systematic measurements of
vector-boson production in association with light jets (see for example
refs.~\cite{Khachatryan:2014uva,Aad:2014qxa,Khachatryan:2016fue}) and these
measurements have shed light on the precision of theoretical and experimental
tools to describe processes with many objects in the final state.
%
%This in turn has been important to gain confidence on the techniques 
% employed for new-physics searches.
%
In the future, it will be interesting to use high-multiplicity processes
including $b$ jets, like the ones studied in this article, to
verify both theory predictions and experimental techniques.
We provided \ntuple{} sets for the predictions obtained by \BlackHat{} and
\SHERPA{} for guidance in future analyses of the 
$Wb\bar b\,\!+\!\,n$-jet
signatures.

\section*{Acknowledgments}
We thank Z.~Bern, L.J.~Dixon, S. H\"oche, D.A.~Kosower, D.~Ma\^itre, L.~Reina and C.~Weiser
for many helpful discussions. 
The work of F.R.A. and F.F.C. is supported by the Alexander von Humboldt
Foundation, in the framework of the Sofja Kovalevskaja Award 2014, endowed by
the German Federal Ministry of Education and Research. 
H.I.'s work is partly supported by the
Juniorprofessor Program of the Ministry of Science, Research and the Arts of the
state of Baden-W\"urttemberg, Germany.
V.S.'s work is funded by the German Research Foundation (DFG) within the
Research Training Group GRK~2044.
This work was performed on the \mbox{bwForCluster} NEMO supported by the state
of Baden-W\"urttemberg through bwHPC and the German Research Foundation (DFG)
through grant no INST 39/963-1 FUGG. 

\FloatBarrier

%%%%%%%%%%%%%%%%%%%%%%%%%%%%%%%%%%%%%%%%%%%%%%%%%%%%%%%%%%%%%%%

%\bibliography{citations}
\bibliography{ref}

\end{document}